\documentclass[aps,pra,twocolumn,superscriptaddress]{revtex4-1}

\usepackage[utf8]{inputenc} 
\usepackage[T1]{fontenc} 
\usepackage[english]{babel}
\usepackage[autostyle=true]{csquotes} 
\usepackage{xcolor}
\usepackage{graphicx}
\usepackage{array}
\usepackage{amsmath} 
\usepackage{amssymb} 
\usepackage{amsfonts} 
\usepackage{amsbsy}
\usepackage{mathrsfs} 
\usepackage{dsfont} 
\usepackage{dcolumn} 
\usepackage{upgreek} 
\usepackage{bm} 
\usepackage{bbm}
\usepackage{placeins} 
\usepackage{physics} 
\usepackage[colorlinks,linkcolor=blue,citecolor=blue,urlcolor=blue]{hyperref}
\usepackage[most]{tcolorbox}
\usepackage{subfigure}
\usepackage{float}
\usepackage[sort&compress]{natbib}
\usepackage[dvips]{epsfig}
\usepackage{hhline}
\usepackage{multirow}
\usepackage[normalem]{ulem}
\newcommand{\amend}[1]{\textcolor{red}{#1}}
\newcommand{\remove}[1]{\textcolor{red}{\xout{(#1)}}}
\renewcommand{\amend}[1]{#1}
\renewcommand{\remove}[1]{}

\def\ai	        {{\em ab--initio}}

\renewcommand{\[}{\left[}
\renewcommand{\]}{\right]}
\renewcommand{\(}{\left(}
\renewcommand{\)}{\right)}
\def\nl         {\right.\\ \left.}

\def\dg         {\dagger}
\def\rar        {\rightarrow}

\newcommand{\olrar}[1]{\overleftrightarrow{#1}}
\newcommand{\orar}[1]{\overrightarrow{#1}}
\newcommand{\olar}[1]{\overleftarrow{#1}}

\newcommand{\eq}[1]{\begin{align}#1\end{align}}
\newcommand{\ml}[1]{\begin{multline}#1\end{multline}}

\newcommand{\eqg}[1]{\begin{gather}#1\end{gather}}
\newcommand{\seq}[1]{\begin{subequations}#1\end{subequations}}
\newcommand{\sst}[2]{\substack{#1\\#2}}
\newcommand{\seql}[2]{\begin{subequations}\label{#1}#2\end{subequations}}
\newcommand{\mll}[2]{\begin{multline}\label{#1}#2\end{multline}}
\newcommand{\eql}[2]{\begin{align}\label{#1}#2\end{align}}
\newcommand{\eqgl}[2]{\seq{\label{#1}\begin{gather}#2\end{gather}}}
\newcommand{\stkout}[1]{\ifmmode\text{\sout{\ensuremath{#1}}}\else\sout{#1}\fi}
\newcommand{\average}[1]{\left\langle #1 \right\rangle}
\newcommand{\phmod}[1]{\left| #1 \right|}

\newcommand{\bpr}[1]{\biggl( #1 \biggr)}
\newcommand{\bpg}[1]{\biggl\{ #1 \biggr\}}
\newcommand{\lab}[1]{\label{#1}}

\newcommand{\ul}[1]{\underline{#1}}
\newcommand{\mat}[1]{\underline{\underline{#1}}}
\newcommand{\oo}[1]{\overline{#1}}
\newcommand{\e}[1]{Eq.~\eqref{#1}}
\newcommand{\es}[2]{Eqs.~\eqref{#1}--\eqref{#2}}
\newcommand{\elab}[2]{Eq.(\ref{#1}#2)}

\newcommand{\fig}[1]{Fig.\ref{#1}}
\newcommand{\figlab}[2]{Fig.\ref{#1}#2}
\newcommand{\evalat}[2]{\left.#1\right|_{#2}}
\renewcommand{\sec}[1]{Section\,\ref{#1}}
\newcommand{\app}[1]{Appendix\,\ref{#1}}
\newcommand{\h}[1]{\hat{#1}}

\newcommand{\ocite}[1]{Ref.\cite{#1}}
\newcommand{\myref}[1]{}
\def\grad{\mbox{\boldmath $\nabla$}}

\def\Re{{\rm Re}}
\def\Im{{\rm Im}}
\newcommand{\p}{\prime}           

\def\ga         {\alpha}
\def\gb         {\beta}
\def\gc         {\gamma}
\def\gC         {\Gamma}
\def\gd         {\delta}
\def\gD         {\Delta}
\def\gee        {\epsilon}
\def\gl         {\lambda}

\def\go         {\omega}
\def\gO         {\Omega}
\def\gr         {\rho}

\def\zero	{{\mathbf 0}}

\def\yy		{{\mathbf y}}

\def\RR		{{\mathbf R}}

\def\xx		{{\mathbf x}}

\def\PP		{{\mathbf P}}

\def\II		{{\mathbf I}}

\def\kk		{{\mathbf k}}
\def\qq		{{\mathbf q}}

\def\Gu         {{\mathbf \tau}}
\newcommand{\di}{\mathrm{d}}
\newcommand{\im}{\mathrm{i}}
\def\bgc{\mbox{\boldmath $\xi$}}

\def\bgt{\mbox{\boldmath $\tau$}}

\def\callF{\mbox{$\mathcal{F}$}}
\def\callG{\mbox{$\mathcal{G}$}}

\def\callP{\mbox{$\mathcal{P}$}}

\def\callT{\mbox{$\mathcal{T}$}}

\newcommand{\cnrism} {Istituto di Struttura della Materia and Division of Ultrafast Processes in Materials (FLASHit) of the National Research Council, via Salaria Km 29.3, I-00016 Monterotondo Stazione, Italy}

\usepackage{multirow}

\usepackage{tikz-feynman}
\newcommand*{\phprop}{
\begin{tikzpicture}
\begin{feynman}
\vertex (a);
\vertex[left= 1.0 cm of a] (b);
\diagram*{(a) --[gluon] (b),};
\end{feynman}%
\end{tikzpicture}%
}
\newcommand*{\Wprop}{
\begin{tikzpicture}
\begin{feynman}
\vertex (a);
\vertex[left= 1.0 cm of a] (b);
\diagram*{(a) --[photon] (b),};
\end{feynman}%
\end{tikzpicture}%
}

\begin{document}

\title{Dynamical electron--phonon vertex correction}
\author{Andrea Marini}
\affiliation{\cnrism}
\begin{abstract}
The dynamical screening of the electron--phonon vertex is caused by the retarded oscillations of the electronic charge following the electron--hole scattering
with a phonon mode. This retardation induces a frequency dependence of the electron--phonon interaction.
Model Hamiltonians and \ai\, approaches have instilled the idea that this retardation is, in most of the cases, negligible.
In this work I demonstrate that the dynamical screening of the electron--phonon vertex cannot be neglected {\em a priori}.
By using a perturbative expansion I introduce a controllable and physically sound method to evaluate and include dynamical screening effects.
Based on the exact results of the homogeneous electron gas I propose a dynamical vertex correction function $\gC_{e-p}\(\go\)$
designed to screen the commonly used adiabatic electron--phonon interaction. This function is expressed in terms of adiabatic quantities, that can be
easily calculated and used to evaluate the strength of the dynamical corrections, even in realistic materials.
\end{abstract}
\date{\today}
\maketitle

\section{Introduction}\lab{sec:intro}
The electron--phonon\,(e--p) vertex is ubiquitous in many different areas of applied and theoretical physics. It appears in both the 
electronic and phononic self--energies~\cite{Stefanucci2023} and its accurate description is crucial in a  wealth of different phenomena. These range from
superconductivity, thermal transport, Raman scattering, phonon mediated absorption and luminescence and also out--of--equilibrium phenomena, just to cite 
few\footnote{More details can be found in textbooks\cite{mattuck1992a,Schrieffer1999,Grimvall1981,ALEXANDERL.FETTER1971} and comprehensive
reviews\cite{Giustino2017}}.

Despite its wide spread application a formal and comprehensive theory of the interaction between electrons and phonons appeared only in 2023~\cite{Stefanucci2023}. 
The work of Stefanucci, Van Leeuwen and Perfetto takes inspiration from a long series of works that, starting from 1961, shed light on different aspects of the problem. 
However, even if we have a formally exact theoretical scheme to tackle the e--p problem, approximations are needed to make realistic
calculations possible. And still in 2023 J. Berges et al.~\cite{Berges2023} discussed at length the screening of the e--p vertex with the
work {\em Phonon Self-Energy Corrections: To Screen, or Not to Screen}. In this work the authors refer to the problem of how to screen of the e--p vertex as a
{\em recently revived controversy}. 

The reason for this very active research activity on the basic properties of the e--p interaction can be understood by looking at the historical 
development. This is well described in \ocite{Stefanucci2023,Marini2023,Giustino2017} and it makes clear that two
main theoretical achievements, model Hamiltonians and Density Functional Perturbation Theory\,(DFPT)\cite{Baroni1987,Gonze1995a,Gonze1995b}, have {\em
de--facto} imposed one of the most stringent and widely accepted approximation currently used in standard calculations: the neglection of retardation effects in
the e--p vertex. The work of Berges~\cite{Berges2023} uses, indeed, DFPT to advocate the adiabatic ansatz where retardation effects are neglected.

The problem of the screening of the e--p vertex can be easily understood by observing that the Hamiltonian which describes the e--p interaction is
\eql{eq:intro.0}
{
 \h{H}\sim \h{H}_0 +\sum_{ij \nu}\evalat{g^{\nu}_{ij}}{bare} \h{c}^\dag_i \h{c}_j\(\h{b}_\nu+\h{b}^\dag_\nu\) +\h{H}_{e-e}.
}
I will define in detail the different terms entering \e{eq:intro.0} later in this work. What is relevant here is that the Hamiltonian must we written in terms
of bare electron--electron\,(e--e)\,($\h{H}_{e-e}$) and bare e--p interaction\,($\h{H}_{e-p}$).  
$\h{H}_0$ is the reference one--body electronic and phononic Hamiltonian.
$\h{H}_{e-e}$ is the  real bottleneck in the practical application of \e{eq:intro.0} as it makes it difficult to solve it by using a single--particle
strategy. At the same time $\h{H}_{e-e}$ plays a crucial role in dressing, mainly via charge oscillations, all elemental processes caused by $\h{H}$. An example
is the popular $GW$~\cite{Onida2001} approximation for the electronic self--energy. $\h{H}_{e-e}$ is written in terms of the bare e--e bare interaction\,($V$),
while in the $GW$ approximation it appears the dynamical e--e interaction $W\(\go\)=\gee^{-1}\(\go\)V$, obtained by screening $V$ via the inverse
dielectric function $\gee^{-1}\(\go\)$. 

The dynamical screening that appears in the $GW$ electronic self--energy is motivated by the fact that the elemental annihilation (or creation) of an
electron--hole pair generates a time--dependent charge oscillation. The very same effect is, in contrast, neglected in the  e--p case.  Model Hamiltonians, like
Fr\"ohlich~\cite{Frohlich1954,Langreth1964,Engelsberg1963,Mahan1990}, are  indeed based on the assumption that \e{eq:intro.0} can be replaced with
\eql{eq:intro.1}
{
 \h{H}_{SCR}\sim \h{H}_0 +\sum_{ij \nu}\evalat{g^{\nu}_{ij}}{SCR} \h{c}^\dag_i \h{c}_j\(\h{b}_\nu+\h{b}^\dag_\nu\).
}
In \e{eq:intro.1}  $\evalat{g^{\nu}_{ij}}{SCR}=\[ \gee^{-1}\(\go=0\)\evalat{g^{\nu}}{bare}\]_{ij}$ is the statically screened e--p interaction.
Electron--electron retardation effects are ignored in \e{eq:intro.1}.  To motivate the different treatment of the electronic screening in the e--p case,
compared to the e--e case, we can argue that, in general, plasma frequencies (corresponding to the poles of $\gee^{-1}\(\go\)$)  are much larger compared to
phonon energies. But this is not always strictly true, especially if we consider metallic systems.
\amend{Indeed the hybridization of plasmons and phonons has been recently observed in doped semiconductors~\cite{Lihm2024}. The same mechanism has been shown,
in \ocite{Krsnik2024}, to cause the statically screened approximation to largely overestimate the superconductive transition temperature in two--dimensional system.}

\amend{Morever, this year, Stefanucci and Perfetto\cite{Stefanucci2025} have clarified some formal aspects linked to the use of \e{eq:intro.1}. In their work they
have demonstrated that it is, indeed, possible to write the {\em imaginary part} of the phonon self--energy solely in terms of two dynamically screened e--p
vertexes. Their result, as explictily stated in \ocite{Stefanucci2025}, does not justify the unconditional of \e{eq:intro.1}.}

It is, however, clear that $\h{H}_{SCR}$ is computationally more simple compared to
$\h{H}$.
In addition to its computational simplicity $\h{H}_{SCR}$ is widely used also for its close link with DFPT, the \ai\, method to calculate phonon properties.
Indeed within DFPT the atomic dynamical matrix is written in terms of $\evalat{g^{\nu}_{ij}}{SCR}$ and $\evalat{g^{\nu}_{ij}}{bare}$.
The close resemblance between the DFPT dynamical matrix and the Many--Body Perturbation
Theory\,(MBPT) phonon self--energy has suggested some authors\cite{Calandra2010,Berges2023,Caldarelli2025} to extend the variational properties of DFPT to the
phonon self--energy to motivate the use of a statically screened e--p vertex.  

\begin{figure}[H]
  {\centering
  \includegraphics[width=\columnwidth]{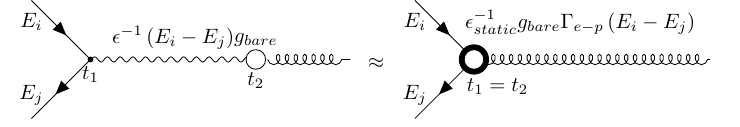}
  }
  \caption{
\footnotesize{ 
Elemental scattering of an electron--hole pair (energy $E_i$ and $E_j$) into a phonon mode. The oscillation induced by
this elemental scattering process delays the creation of the phonon. In this paper I demonstrate that the simple form 
described by \es{eq:intro.1}{eq:intro.2} can be, actually, used at the price of introducing a frequency dependent vertex correction $\gC_{e-p}\(E_i-E_j\)$.
}
\label{fig:1}
}
\end{figure}

If model Hamiltonians and DFPT have largely promoted the use of \e{eq:intro.1}, also from a physical point of view, 
$\h{H}_{SCR}$ may appear to be physically sound and intuitive. Indeed when $\h{H}_{SCR}$ is treated within perturbation
theory~\cite{Sakurai1994} the probability for the scattering of the electron from state $i$ to state $j$ by emitting/absorbing a phonon of frequency
$\go_\nu$ is found to be
\eql{eq:intro.2}
{
 \callP_{i\rar j} \sim \sum_{\nu} | \evalat{g^{\nu}_{ij}}{SCR} |^2 \gd\(E_i-E_j\pm\go_\nu\).
}
\e{eq:intro.2} is at the basis of countless applications of the electron--phonon theory~\cite{LafuenteBartolome2022,Sadasivam2017,Allen1987,Giustino2017}.
If, however, we analyze the role of retardation starting from \e{eq:intro.0} and using MBPT we readily see that the microscopic, elemental phonon scattering
process is graphically represented in the left frame of \fig{fig:1}.  An electron--hole pair is annihilated at $t=t_1$ into a charge, collective, oscillation
(\Wprop). This elemental excitation delays the creation of a phonon (\phprop) of $\Delta t=t_2-t_1$. If relevant, this delay would make impossible to rewrite
the scattering probability as \e{eq:intro.2} and described by \e{eq:intro.1}.

Still it would be desirable to embody all dynamical corrections in a unique, frequency dependent, function $\gC_{e-p}\(\go\)$ able to amend the statically
screened vertex with a correction dependent on the electron--hole pairs energy involved in the process. This would the most natural extension
of \e{eq:intro.1} that, even keeping the simple Fermi golden--rule like form, fully includes dynamical corrections. 

This is the main goal of this work, where 
I investigate, within MBPT, the role of dynamical screening in the definition of the e--p vertex. The goal is to base on a solid mathematical ground
the conditions under which \e{eq:intro.1} can be used.

I start in \sec{sec:H} by reviewing the different ingredients of 
the Hamiltonian: bare phonons, bare electrons and the electron--electron and
the e--p interactions. The e--p vertex is introduced and discussed in \sec{sec:vertex_and_SEs} together with the definition of the phonon self--energy on the
Keldysh contour. The equilibrium limit is defined in \sec{sec:eq_limit}. In \sec{sec:Pi_and_QPH} I review the quasi--phonon solution of the Dyson equation. 

In \sec{sec:SSA} I discuss how dynamical effects naturally emerge from a mathematical inspection of the variational arguments of
\ocite{Berges2023,Calandra2010,Caldarelli2025}.
In \sec{sec:G_dyn} I discuss how to embody the dynamical screening of the e--p vertex, taking as reference the adiabatic limit, in a fully coherent
MBPT scheme. I derive an expression for a vertex dynamical correction $\gC_{e-p}\(\go\)$ and\amend{, inspired by the seminal
works\cite{Falter1988,Falter1981,Falter1980} of C.Falter} I introduce its
dynamical perturbative expansion\,(DPE). I demonstrate that the widely used doubly statically screened approximation for the phonon self--energy corresponds to 
the first order in the DPE, thus defining the range of validity for \es{eq:intro.1}{eq:intro.2} to be correct.
I explicitly calculate higher order terms of the DPE of $\gC_{e-p}\(\go\)$, corresponding to an increasing strength of the dynamical effects. 

In \sec{sec:jells} I move to the validation of the theory in a generalized, periodic, three--dimensional electron gas\,(PHEG). 
In \sec{sec:JELL_QPH} I use the PHEG to discuss the validity of the commonly used one--the--mass shell approximation\,(OMS) compared to the exact
solution and the  QPHA. In \sec{sec:JELL_Pi} and \sec{sec:JELL_Gamma} I calculate and discuss the different orders in the PDE of 
$\gC_{e-p}\(\go\)$ and of the phonon self--energy. 
I discuss how the PHEG represents a stringent test on the validity of the static screening approximation due to the presence of a single plasmon peak
that, in some cases, approaches the phonon frequency. When the two frequencies are comparable the Taylor expansion diverges and dynamical corrections become
non perturbative.

I also provide a simple form of $\gC_{e-p}\(\go\)$ written in terms of bare and statically screened e--p vertexes, that can be used to estimate the
vertex correction function even in realistic materials. 

The final message is that dynamical corrections are, in general, not negligible and, depending on the presence of resonances with energies comparable to the
phonon frequencies, can require a non perturbative, fully dynamic treatment, well beyond the 1$^{st}$ order. This will confirm that the
\es{eq:intro.1}{eq:intro.2} are not well motivated, in general, and great care must be used in order to numerically simulate realistic materials.

\section{The Hamiltonian}\lab{sec:H}
I consider here a collection of $N$ identical atoms, defined in terms of quantistic positions $\{\h{\RR}_a\}$, momenta $\{\h{\PP}_a\}$, nuclear charge $Z$ and
mass $M$\footnote{For simplicity I consider one atomic species. The extension to the more general case is straightforward and can be found, among others, in
\ocite{Marini2023,Marini2015}.}. The atoms are arranged at the corners of a periodic lattice whose elemental cell volume is $V_0$. The lattice volume is,
instead, $V_c=N V_0$.  I follow here the derivation presented in detail in \ocite{Stefanucci2023,Marini2023,Marini2015,Mahan1990}.  Bold variables represent
Cartesian vectors and I used latin letters ($a,b,...$) to denote the generic atom. 

The total Hamiltonian of such a system is
\eql{eq:H.1}
{
 \h{H}=\(\h{T}_n+\h{H}_{n-n}\)+\(\h{H}_e+\h{H}_{e-e}\)+\h{H}_{e-n}.
}
In \e{eq:H.1} I have introduced: the bare electrons (nuclei) part $\h{H}_{e\(n\)}$,  the electron--nuclei interaction $\h{H}_{e-n}$, the electron--electron term
$\h{H}_{e-e}$, the nuclear kinetic operator $\h{T}_n$ and the nucleus--nucleus interaction, $\h{H}_{n-n}$.  I use $\h{O}$ to indicate an operator.  
It is important to remind that 
$\h{T}_n$, $\h{H}_{n-n}$ and $\h{H}_{e-n}$ depend on the quantized atomic positions ($\{\h{\RR}_a\}$) and momenta ($\{\h{\PP}_a\}$).

In order to introduce phonons, as bosonic operators, let's introduce reference, equilibrium
atomic positions ($\{\RR^0_a\}$), and quantized displacements ($\{\h{\Gu}_a\}$):
\eql{eq:H.2}
{
 \h{\RR}_a=\RR^0_a+\h{\bgt}_a.
}
Thanks to \e{eq:H.2} we can perform a formal Taylor expansion of \e{eq:H.1} in powers of $\h{\bgt}_a$. By making explicit the dependence of $\h{H}$ from the
atomic position operators we have
\mll{eq:H.2.1}
{
 \h{H}_{e/n-n}\(\{\h{\RR}_a\}\)\sim 
 \h{H}_{e/n-n}\(\{\RR^0_a\}\)+\\+\sum_a \grad_{\RR_a} H_{e/n-n}\(\{\RR^0_a\}\) \cdot \h{\bgt}_a.
}
\e{eq:H.2.1} applies to both $\h{H}_{e-n}$ and $\h{H}_{n-n}$.

Following \ocite{Marini2015,Marini2023} I add and remove from \e{eq:H.1} $\h{H}_{ref}$, that is the e--n contribution to the
dynamical matrix of a reference Born--Oppenheimer\,(BO) surface used to define the equilibrium positions:
\eql{eq:H.2.2}
{
 \h{H}_{ref}\(\{\RR^0_a\}\)=\frac{1}{2}\sum_{ab}\nolimits^\p  \h{\bgt}_a \cdot \olrar{C}^{ref}_{ab} \cdot  \h{\bgt}_b.
}
\e{eq:H.2.2} plays the same role of the reference electronic density, $n^0\(\xx\)$, introduced in \ocite{Stefanucci2023}.

We now introduce the reference
single--particle electronic basis defined as eigenstates of $\h{H}_e+\h{H}_{e-n}\(\{\RR^0_a\}\)$. These states define the basis to introduce the field operators
$\h{\psi}_{n\kk}\(\xx\)=\phi_{n\kk}\(\xx\) \h{c}_{n\kk}$, with $\h{c}_{n\kk}$ the single--particle annihilation operator of an electron
on the level $n$ with momentum $\kk$ and  energy $\gee_{n \kk}$.
$\phi_{n\kk}\(\xx\)=e^{\im \kk\cdot \xx} u_{n\kk}\(\xx\)$ the corresponding wave--function, $u_{n\kk}\(\xx\)$ the periodic part.  
In this basis we have that
\eql{eq:H.3}
{
 \h{H}_e+\h{H}_{e-n}\(\{\RR^0_a\}\)=\sum_{n \kk} \gee_{n \kk}  \h{c}^\dg_{n\kk} \h{c}_{n\kk}.
}

Similarly we can now expand in powers of $\h{\bgt}_a$ the nuclear contributions to $\h{H}$. 
The final result is that
\mll{eq:H.6}
{
 \h{T}_n+\h{H}_{n-n}\(\{\h{\RR}_a\}\)+\h{H}_{ref}\(\{\RR^0_a\}\)=\\=E^0_{BO}+\frac{1}{2}\sum_{\qq\gl}\go_{\qq\gl}\(2\h{n}_{\qq\gl}+1\),
}
with $E^0_{BO}$ the equilibrium reference BO energy.
$\qq$ and $\gl$ are phonon momentum and branch introduced via the standard canonical transformation
\eql{eq:H.7}
{
 \h{\bgt}_a=
-\frac{\im}{\sqrt{NM}}
\sum_{\qq\gl}
\frac{1}{\sqrt{\go_{\qq\gl}}}
{\bgc}_{\qq\gl}\h{u}_{\qq\gl}e^{\im \qq\cdot \RR^0_a},
}
with the polarization vectors $\ul{\xi}_{\qq\gl}$ and
\eql{eq:H.8}
{
 \h{u}_{\qq\gl}=\frac{1}{\sqrt{2}}\(\h{b}_{\qq\gl}+\h{b}^\dg_{-\qq\gl}\).
}
In \e{eq:H.8} $\h{b}_{\qq\gl}$ and $\h{b}^\dg_{\qq\gl}$ are annihilation and creation phonon operators. In \e{eq:H.6} $\h{n}_{\qq\gl}=\h{b}^\dg_{\qq\gl}
\h{b}_{\qq\gl}$.

\begin{widetext}
By using again \e{eq:H.2.1} we can introduce the linear electron--phonon interaction:
\seql{eq:H.9}{
\ml{
 \h{H}_{e-p}=\h{H}_{e-n}-\h{H}_{e-n}\(\{\RR_0^a\}\)+
  \sum_a \grad_{\RR_a} H_{n-n}\(\{\RR^0_a\}\) \cdot \h{\bgt}_a=\\
  \sum_{\qq\gl} \int_0 g_{\qq\gl}\(\xx\)  \h{\gr}_{-\qq}\(\xx\) \h{u}_{\qq\gl}  e^{-\im \qq\cdot \xx}+ \sum_a \grad_{\RR_a} H_{n-n}\(\{\RR^0_a\}\) \cdot \h{\bgt}_a,
}
where
\eq{
 g_{\qq\gl}\(\xx\)=\sum_{a}\frac{1}{\sqrt{NM\go_{\qq\gl}}} \grad_{\RR_a} V_{e-n}\(\xx-\RR^0_a\)
 \cdot {\bgc}_{\qq\gl}e^{\im \qq\cdot \RR^0_a}
}
}
\end{widetext}
In \e{eq:H.9} I have assumed that 
\eql{eq:H.10}{
\h{H}_{e-n}\(\{\h{\RR}_a\}\)=\sum_a \int_c \di \xx V_{e-n}\(\xx-\h{\RR}_a\) \h{\gr}\(\xx\),
}
and I have introduced the density operator
\mll{eq:H.11}{
\h{\gr}\(\xx\)=\frac{1}{N}\sum_{nm\kk\qq} \oo{u_{n\kk-\qq}\(\xx\)} u_{m\kk}\(\xx\) \h{c}^\dag_{n\kk-\qq} \h{c}_{m\kk} e^{\im\qq\cdot\xx}=\\
\frac{1}{N}\sum_\qq \h{\gr}_\qq\(\xx\)e^{\im\qq\cdot\xx}.
} 
It is important to note that the integral $\int_c\dots$ in \e{eq:H.10} is extended over the entire crystal, while the integral $\int_0\dots$ in
\elab{eq:H.9}{a} is restricted to the unit cell. 

An other important ingredient of the theory are the equilibrium atomic positions $\{\RR^0_a\}$. As discussed in \ocite{Marini2015} those are formally defined by the condition
that the atomic classical forces defined on the BO surface are zero. This defines
$\gD\h{\gr}_{\qq}\(\xx\)=\h{\gr}_{\qq}\(\xx\)-\average{\h{\gr}_{\qq}\(\xx\)}_{BO}$ and \elab{eq:H.9}{a} reduces to
\eql{eq:H.11.1}{
 \h{H}_{e-p}=
  \sum_{\qq\gl} \int_0 g_{\qq\gl}\(\xx\)  \gD\h{\gr}_{-\qq}\(\xx\) \h{u}_{\qq\gl}  e^{-\im \qq\cdot \xx}.
}

The e--e interaction can be similarly worked out obtaining
\eql{eq:H.12}
{
 \h{H}_{e-e}=\frac{1}{2}\sum_\qq \int_0 \di \xx \yy \h{\gr}_{\qq}\(\xx\) V_\qq\(\xx-\yy\)\h{\gr}_{-\qq}\(\yy\),
}
where $V_\qq\(\xx\)$ is defined by the discrete Fourier transform\cite{AMH,Ashcroft1976}
\eql{eq:H.13}{
 \frac{1}{|\xx|}=\sum_\qq V_\qq\(\xx\) e^{\im\qq\cdot\xx}.
}

\section{Dressing of the electron--phonon interaction and the phonon self--energy}\lab{sec:vertex_and_SEs}
The electron--phonon vertex introduced in \e{eq:H.9} is bare, meaning that it is not dressed by the electronic correlations. By means of standard MBPT
techniques~\cite{Stefanucci2023,mattuck1992a,Schrieffer1999,Grimvall1981,Marini2023,Marini2015,Mahan1990,ALEXANDERL.FETTER1971,Leeuwen2004a} it is known that the electron--electron interaction
dresses the electron--phonon vertex in a subtle way. While the electronic self--energy is written in terms of solely screened
vertexes, in the exact phonon self--energy a combination of screened and bare vertexes appear.

The appearance of a bare vertex  in the phonon self--energy is, actually, the consequence of the partition of the total electron--electron interaction in  the
sum of a purely electronic and and a phonon--mediated components. 
If  $W^{total}_{\qq}\(\xx_1 z_1,\xx_2 z_2\)$ is the total electron--electron interaction between points $\(\xx_1 z_1\)$ and $\(\xx_2 z_2\)$
the final result is 
\seql{eq:DP.1}{
\eqg
{
 W^{total}_{\qq}\(1,2\)=W^{p}_{\qq}\(1,2\)+W^{e}_{\qq}\(1,2\),\\
 W^{e}_{\qq}\(1,2\)=\gee_\qq^{-1}\(1,3\)V_\qq\(3,2\),
}
and
\eq{
 W^{p}_{\qq}\(1,2\)=\sum_{\gl_1 \gl_2} \callG_{\qq\gl_1}\(1,z_3\)D_{\qq\gl_1\gl_2}\(z_3,z_4\) \callG_{\qq\gl_2}\(z_4,2\).
}
}
In \e{eq:DP.1} time is introduced and represented together with the space variable, $1=\(\xx_1,z_1\)$. I also assume 
repeated indexes to be summed or integrated (depending on their definition).
Time is assumed to be on the Keldysh contour~\cite{Leeuwen2013} so that the general theory applies both at equilibrium and out--of--equilibrium. 
Moreover I introduce $V_\qq\(1,2\)=\gd\(z_1-z_2\) V_\qq\(\xx_1-\xx_2\)$.
As discussed in \ocite{Marini2023} in the case of the phonon self--energy is essential, in order to derive sound approximations, to obtain the equilibrium phonon
self--energy from the general expression on the Keldysh contour.

The partition defined by \e{eq:DP.1} defines the left and right Dyson equations for the phonon propagator
$\mat{D}$
\begin{widetext}
\seql{eq:DP.4.2}{
\eq{
 D_{\qq\gl_1\gl_2}\(z_1,z_2\)=D^{0}_{\qq\gl_1}\(z_1,z_3\)\bpg{\gd_{32}\gd_{\gl_1\gl_2} +
 \[ \orar{\Pi}_{\qq\gl_1\gl_3}\(z_3,z_4\)-C^{ref}_{\qq\gl_1\gl_3}\gd\(z_3-z_4\)\]D_{\qq\gl_3\gl_2}\(z_4,z_2\)},
}
and
\eq{
 D_{\qq\gl_1\gl_2}\(z_1,z_2\)=\bpg{\gd_{\gl_1\gl_2}\gd_{13} +
 D_{\qq\gl_1\gl_3}\(z_1,z_4\) \[\olar{\Pi}_{\qq\gl_3\gl_2}\(z_4,z_3\)-C^{ref}_{\qq\gl_3\gl_2}\gd\(z_3-z_4\)\]}D^{0}_{\qq\gl_2}\(z_3,z_2\).
}}
\end{widetext}
In \e{eq:DP.4.2} we have introduced the representation of $\olrar{C}_{ab}$ in the phonons basis, $C^{ref}_{\qq\gl_3\gl_2}$, and the
left and right out--of--equilibrium phonon self--energies
\eqgl{eq:DP.4.3}{
 \orar{\Pi}_{\qq\gl_1\gl_2}\(z_1,z_2\)= g_{-\qq\gl_1}\(\xx_1\)  \chi^0_\qq\(1,3\)  \callG_{\qq\gl_2}\(3,z_2\),\\
 \olar{\Pi}_{\qq\gl_1\gl_2}\(z_1,z_2\)= \callG_{-\qq\gl_1}\(z_1,3\) \chi^0_\qq\(3,2\)  g_{\qq\gl_2}\(\xx_3\),
}

The dressed electron--phonon vertexes that enter in \e{eq:DP.4.3} are
\eqgl{eq:DP.5}{
\callG_{\qq\gl_2}\(1,z_2\)= \gee^{-1}_{\qq}\(1,2\) g_{\qq\gl_2}\(\xx_2\).\\
\callG_{-\qq\gl_1}\(z_1,2\)= g_{-\qq\gl_1}\(\xx_1\) \gee^{-1}_{\qq}\(1,2\).
}

\subsection{Equilibrium limit}\lab{sec:eq_limit}
In \ocite{Marini2023} it has been demonstrated that the equilibrium limit of the phonon self--energy is a combination of the left and right self--energies.

In this limit the retarded phonon propagator satisfies the Dyson equation
\seql{eq:eq.1}
{
\ml
{
D_{\qq\gl_1\gl_2}\(t\)=D^{0}_{\qq\gl_1}\(t-t^\p\)\bpr{\gd_{\gl_1\gl_2}\gd\(t-t^\p\)
+\\+\[\Pi_{\qq\gl_1\gl_3}\(t^\p-\oo{t}\)-C^{ref}_{\qq\gl_1\gl_3}\gd\(t^\p-\oo{t}\)\]D_{\qq\gl_3\gl_2}\(\oo{t}\)},
}
with
\eq
{
\Pi_{\qq\gl_1\gl_2}\(t\)=\frac{1}{2}\[ \orar{\Pi}_{\qq\gl_1\gl_2}\(t\)+\olar{\Pi}_{\qq\gl_1\gl_2}\(t\)\].
}
}
By taking the equilibrium limit of $\callG_{\qq\gl}$ it follows that
\mll{eq:eq.2}
{
\Pi_{\qq\gl_1\gl_2}\(t_1-t_2\)=\\ \frac{1}{2}\[ 
g_{-\qq\gl_1}\(\xx_1\)  \chi^0_\qq\(\xx_1 \xx_3,t_1-t_3\)  \callG_{\qq\gl_2}\(\xx_3, t_3-t_2\)+\nl+
\callG_{-\qq\gl_1}\(t_1-t_3,\xx_3\)  \chi^0_\qq\(\xx_3 \xx_2,t_3-t_2\)  g_{\qq\gl_2}\(\xx_2\)
\]
}
\e{eq:eq.2} is a very powerful definition of the phonon self--energy. Indeed if $\callG_{\qq\gl_2}\(\xx t\)$
is calculated exactly the left and right self--energies are identical and \e{eq:eq.2} reduces to one term. At the same time, however,
it gives us the freedom to approximate $\ul{\callG}_{\qq}$ in such a way to break the symmetry of the individual left/right components still ensuring a fully
symmetric total self--energy.

The last ingredient we need to define explicitly is the inverse dynamical dielectric function $\gee^{-1}_{\qq}\(1,2\)$ that appears in \e{eq:DP.1}. This is done 
in \app{APP:TDH} by introducing a combined index for bands pairs and momenta $\II_i\equiv\(n_i,m_i,\kk_i\)$. 

Any non--local operator $O\(1,2\)$ can be represented in this basis by
using 
\mll{eq:eq.2.1}{
  O_{\II_1 \oo{\II}_2}\(t_1,t_2\)=\int_0 \di \xx_1\xx_2  \oo{u_{n_1\kk_1-\qq}\(\xx_1\)} u_{m_1\kk_1}\(\xx_1\)\\
 \oo{u_{n_2\kk_2}\(\xx_2\)} u_{m_2\kk_2-\qq}\(\xx_2\) O\(\xx_1t_1,\xx_2 t_2\).
}
The $\oo{\II}_2$ in \e{eq:eq.2.1} appears to keep track of the change of order in the left and right pair of wave--functions.
It is simple algebra to demonstrate that, in the $\{\II_i\}$ basis, the Fourier transformed of the equilibrium  $\gee^{-1}_{\qq}\(1,2\)$ is
\eql{eq:eq.3}{
 \mat{\gee}_\qq^{-1}\(\go\)=\mat{1}+\mat{V}_\qq \mat{\chi}_\qq\(\go\),
}
with $\mat{\chi}$ the electronic response function matrix that solves the time--dependent Hartree equation
\eql{eq:eq.5}
{
\mat{\chi}_\qq\(\go\)=\mat{\chi}^0_\qq\(\go\)\[\mat{1}+\mat{V}_\qq\mat{\chi}_\qq\(\go\)\].
}  
Now, as both $\mat{\chi}$ and $\mat{\Pi}$ are matrices, in order to keep the formalism  simple in view of the rest of the paper, I assume the phonon
self--energy and propagator to be diagonal
in the branch index. Thanks to this assumption we can transform the spatial integration's in \e{eq:eq.2} into matrix multiplications:
\mll{eq:eq.6}
{
 \Pi_{\qq\gl}\(\go\)=\frac{1}{2}\sum_{\II_1\II_2}\[ g_{-\qq\gl,\II_1}  \chi^0_{\qq\II_1\oo{\II}_2}\(\go\)
\callG_{\qq\gl,\oo{\II}_2}\(\go\) + \nl+  \callG_{-\qq\gl,\II_1}\(\go\)  \chi^0_{\qq\II_1\oo{\II}_2}\(\go\)
g_{\qq\gl,\oo{\II}_2} \]=\\
 \frac{1}{2}\[ \ul{g}_{-\qq\gl}^T \mat{\chi}^0_\qq\(\go\) \ul{\callG}_{\qq\gl}\(\go\)+\ul{\callG}_{-\qq\gl}^T\(\go\) \mat{\chi}^0_\qq\(\go\) \ul{g}_{\qq\gl}\].
}
In \e{eq:eq.6} I have introduced a compact notation for \e{eq:DP.5}, noticing that from
\eql{eq:eq.7}{
 \ul{\callG}_{\pm\qq\gl}\(\go\)= \mat{\gee}^{-1}_{\qq}\(\go\) \ul{g}_{\pm\qq\gl},
}
it follows $\ul{\callG}^T_{-\qq\gl}\(\go\)= \ul{g}^T_{-\qq\gl}  \mat{\gee}^{-1,T}_{\qq}\(\go\)$.

\subsection{Quasi--phonon solution}\lab{sec:Pi_and_QPH}
Once we have the phonon self--energy we need a recipe to extract from the Dyson equation the final renormalized phonon width and energy.
In the case of electrons this procedure is known as quasi--particle approximation\,(QPA)~\cite{0034-4885-61-3-002,Onida2002,10.3389/fchem.2019.00377}.

The goal of this section is to review the quasi--phonon approximation\,(QPHA), derived independently in \ocite{Marini2024} and \ocite{Stefanucci2023} that, 
in complete analogy to the electronic quasi--particle, will define both the phonon energy
and width. As it will be clear shortly the QPHA will impose any approximation to describe accurately both the real and imaginary part of the self--energy.

Let's start from the retarded, frequency dependent, independent--particle phonon propagator 
\eql{eq:QPH.1}
{
 D^{0}_{\qq\gl}\(\go\)= \frac{\go_{\qq\gl}}{\(\go+\im 0^+\)^2 - \go_{\qq\gl}^2}.
}
In \e{eq:QPH.1} does not appear the factor $2$ at the numerator as a consequence of the $\sqrt{2}$ appearing in \e{eq:H.7}.
In the diagonal case the Fourier transformed of \elab{eq:eq.1}{a} is
\eql{eq:QPH.1.1}
{
 D_{\qq\gl}\(\go\)= \frac{1}{\(D^{0}_{\qq\gl}\(\go\)\)^{-1}- \Pi_{\qq\gl}\(\go\)+C^{ref}_{\qq\gl}}.
}

At this point I formally introduce a QPHA representation of $\Pi_{\qq\gl}\(\go\)$:
\seql{eq:QPH.2}{
\eq
{
 \evalat{\Pi_{\qq\gl}\(\go\)}{QPHA}= \evalat{\Pi_{\qq\gl}}{s}+ \im \Im\[ \gb_{\qq\gl} \] \go + \Re\[ \gb_{\qq\gl} \] \frac{\go^2}{\go_{\qq\gl}},
}
where
\eq{
 \gb_{\qq\gl}=\frac{\Pi_{\qq\gl}\(\go_{\qq\gl}\)-\evalat{\Pi_{\qq\gl}}{s}}{\go_{\qq\gl}}.
}
}
In \e{eq:QPH.2} I have used $\evalat{\(\cdots\)}{s}$ to represent the adiabatic limit of $\(\cdots\)$. I will use the same notation throughout the work.

\e{eq:QPH.2} is formally very different from the QPA form of the electronic self--energy. Indeed in the QP the self--energy is expanded at the first order while
\e{eq:QPH.2} needs a second order term. The reason of this difference traces back to the different frequency dependence of the self--energies.
The QPA is motivated by the fact that the poles of electron--electron interaction describe plasmonic excitations. In general plasmon excitations are,
energetically, very far from the single--particle levels. When the electron--electron interaction is mediated by phonons this argument does not hold anymore and,
indeed, it is possible to observe large deviations from the QPA~\cite{Cannuccia2011a}.

The case of the phonon self--energy is even more special. The $\Pi_{\qq\gl}\(\go\)$ frequency dependence, in the homogeneous electron gas, is dominated~\cite{Engelsberg1963}
by a fast rise of imaginary part which corresponds to a slower rise of the real part. This is the reason of the two terms in \e{eq:QPH.2}.

If we now use \e{eq:QPH.2} in the \e{eq:QPH.1.1} it is simple math to show that 
\eql{eq:QPH.6.3}
{
 \evalat{D_{\qq\gl}\(\go\)}{QPHA}= \frac{\go_{\qq\gl} \remove{Z_{\qq\gl}}}{\(\go+\im \gamma^{QPHA}_{\qq\gl}\)^2-\(\gO^{QPHA}_{\qq\gl}\)^2},
}
where
\begin{widetext}
\eqgl{eq:QPH.4}
{
 Z_{\qq\gl}=\frac{1}{1-\Re\[\gb_{\qq\gl}\]},\\
 \gamma^{QPHA}_{\qq\gl}=-\frac{Z_{\qq\gl}}{2}\Im\[\Pi_{\qq\gl}\(\go_{\qq\gl}\)\],\\
 \gO^{QPHA}_{\qq\gl}=\pm\sqrt{\[Z_{\qq\gl}\go_{\qq\gl}\(\go_{\qq\gl}+\evalat{\Pi_{\qq\gl}}{s}-C^{ref}_{\qq\gl}\)-\(\gamma^{QPHA}_{\qq\gl}\)^2\]}.
}
\end{widetext}
From \e{eq:QPH.4} we see that in the QPHA it appears a renormalization factor, $Z_{\qq\gl}$, that is the phonon counterpart of the QPA renormalization factor.

The phonon on--the-mass shell\,(OMS)~\cite{Giustino2017} approximation can be obtained from \es{eq:QPH.6.3}{eq:QPH.4} by assuming that $\gb_{\qq\gl} \ll 1$.
In this limit indeed
\seql{eq:QPH.5}{
\eq{
 Z_{\qq\gl}\sim 1+\Re\[\gb_{\qq\gl}\],
}
and
\eq{
 \sqrt{Z_{\qq\gl}}\sim 1+\frac{\Re\[\gb_{\qq\gl}\]}{2}.
}}
From \e{eq:QPH.5} it follows
\seql{eq:QPH.6}{
\eq
{
 \gamma^{QPHA}_{\qq\gl}\approx -\frac{1}{2}\Im\[\Pi_{\qq\gl}\(\go_{\qq\gl}\)\]=\gamma^{OMS}_{\qq\gl},
}
and
\ml
{
 \gO^{QPHA}_{\qq\gl}\approx 
\(1+\frac{\Re\[\gb_{\qq\gl}\]}{2}\)\times\\\times
\sqrt{\(\go_{\qq\gl}+\evalat{\Pi_{\qq\gl}}{s}-C^{ref}_{\qq\gl}\)\go_{\qq\gl}}=
\gO^{OMS}_{\qq\gl}.
}
}

If we  assume $\evalat{\Pi_{\qq\gl}}{s}\approx C^{ref}_{\qq\gl}$ \elab{eq:QPH.4}{c} and \elab{eq:QPH.6}{b} simplify to
\eqgl{eq:QPH.7}
{
 \(\gO^{QPHA}_{\qq\gl}\)^2=Z_{\qq\gl}\go^2_{\qq\gl}-\(\gamma^{QPHA}_{\qq\gl}\)^2,\\
 \gO^{OMS}_{\qq\gl}=\go_{\qq\gl}+\frac{\Re\[\Pi_{\qq\gl}\(\go_{\qq\gl}\)\]-\evalat{\Pi_{\qq\gl}}{s}}{2}.
}

\es{eq:QPH.6}{eq:QPH.7} make also clear that, if $\gb_{\qq\gl}$ is large it is not possible to write the phonon width solely in terms of
$\Im\[\Pi_{\qq\gl}\]$. 
The OMS and QPHA approximations are compared in the case of the Homogeneous Electron gas in \sec{sec:JELL_QPH}.

\section{The statically screened approximation}\lab{sec:SSA}
The actual evaluation of the full frequency dependence of $\callG_{\qq\gl}\(\xx,\go\)$ and $\callG_{\qq\gl}\(\go,\xx\)$ requires an enormous computational cost. The
reason is that the common applications of the e--p vertex require to use very large $\kk$--point grids that make, in practice, numerically challenging to go beyond the
static approximation.

As discussed in the introduction the applications of model Hamiltonians
and the  wide--spread use of DFPT \ai\, methods and codes, 
have favored the use of a statically screened electron--phonon interaction:
\eqgl{eq:SS.1}{
 \evalat{\callG_{-\qq\gl}\(\xx_2\)}{s}=
 \int\di \xx_1 g_{-\qq\gl}\(\xx_1\) \gee^{-1}_{\qq}\(\xx_1 t_1,\xx_2 t_1^+\),\\
 \evalat{\callG_{\qq\gl}\(\xx_1\)}{s}=
\int\di \xx_2\gee^{-1}_{\qq}\(\xx_1 t_1,\xx_2 t_1^+\)
g_{\qq\gl}\(\xx_2\),
}
with $t_1^+=t_1+0^+$.
\e{eq:SS.1} can be rewritten in matrix form by using the notation introduced in \sec{sec:eq_limit}
\eqgl{eq:SS.2}{
 \evalat{\ul{\callG}_{-\qq\gl}}{s}=\lim_{t\rar 0^+} \ul{g}_{-\qq\gl} \mat{\gee}^{-1}_{\qq}\(t\)=\ul{g}_{-\qq\gl} \evalat{\mat{\gee}^{-1}_{\qq}}{s},\\
 \evalat{\ul{\callG}_{\qq\gl}}{s}=\lim_{t\rar 0^+} \mat{\gee}^{-1}_{\qq}\(t\) \ul{g}_{\qq\gl}=\evalat{\mat{\gee}^{-1}_{\qq}}{s} \ul{g}_{\qq\gl}.
}
The discussion of how to use $\evalat{\ul{\callG}_{\qq\gl}}{s}$ neglecting the full dynamical form of the e--p vertex has inspired several of works that,
starting from \ocite{Calandra2010} up to very recent \ocite{Berges2023,Caldarelli2025,Stefanucci2025}, have attempted to provide a rigorous basis to the static
screening assumption. 
These works are based on two different approaches: the variational properties of the phonon self--energy~\cite{Calandra2010,Berges2023,Caldarelli2025} and the 
down--folding from the Keldysh contour to the real time axis~\cite{Stefanucci2025}. 

In order to simplify the notation of this section I consider the homogeneous limit. This is particularly simple to
introduce in the present, momentum resolved, notation. Indeed it corresponds to neglect all microscopic spatial dependencies such that,
for example, in \e{eq:H.9} $g_{\qq\gl}\(\xx\)\rar g_{\qq\gl}$.

In this case \e{eq:eq.6} becomes
\eql{eq:VAR.1}
{
2 \Pi_{\qq\gl}\(\go\)= g_{-\qq\gl}  \chi^0_\qq\(\go\) \callG_{\qq\gl}\(\go\)+ \callG_{-\qq\gl}\(\go\)  \chi^0_\qq\(\go\)  g_{\qq\gl}.
}
I further assume $g_{\qq\gl}\in \mathbb R$ so that $g_{-\qq\gl}=g_{\qq\gl}$ and $\Pi_{\qq\gl}\(\go\)= g_{\qq\gl}  \chi^0_\qq\(\go\) \callG_{\qq\gl}\(\go\)$.

By using \e{eq:VAR.1} it is simple algebra to show that
\eql{eq:SSK.1}
{
 \callG_{\qq\gl}\(\go\)=\frac{g_{\qq\gl}}{1-V_\qq \chi_\qq^{0}\(\go\)}.
}
If now we use \e{eq:SSK.1} to write the phonon self--energy we get
\mll{eq:SSK.2}
{
\Pi_{\qq\gl}\(\go\)=  g_{\qq\gl} \chi_\qq^{0}\(\go\)\callG_{\qq\gl}\(\go\)=\\
g_{\qq\gl}^2\[\frac{\chi_\qq^{0}\(\go\)-V_\qq |\chi_\qq^{0}\(\go\)|^2}{\amend{|1-V_\qq \chi_\qq^{0}\(\go\)|^2}}\].
}
From \e{eq:SSK.2} it easily follows that
\mll{eq:SSK.3}{
 \Im\[\Pi_{\qq\gl}\(\go\)\]=g_{\qq\gl}^2\[\frac{\Im\[\chi_\qq^{0}\(\go\)\]}{\amend{|1-V_\qq \chi_\qq^{0}\(\go\)|^2}}\]=\\=|\callG_{\qq\gl}\(\go\)|^2
\Im\[\chi_\qq^{0}\(\go\)\],
}
\e{eq:SSK.3} is, clearly, an exact rewriting of the {\em imaginary part} of the retarded phonon self--energy. The proof of \e{eq:SSK.1} in the
out--of--equilibrium regime by down--folding the Keldysh expression to the real time axis can be found in \ocite{Leeuwen2013,Stefanucci2025}. 
In \ocite{Stefanucci2025} Stefanucci and Perfetto use \e{eq:SSK.3} to demonstrate that the static limit can be taken without introducing double counting errors.
In \ocite{Marini2023} it was demonstrated, instead, that when the whole self--energy is considered, and not only its imaginary part, the statically screened
approximation can be obtained from a many--body diagrammatic expansion only at the price of including double counting terms.

The two works do not contradict each other as they represent two alternative ways of obtaining the statically screened limit of the {\em imaginary part} of the phonon
self--energy. Still the problem is how to give a precise estimation of the error introduced when we approximate
$\callG_{\qq\gl}\(\go\)\sim\evalat{\callG_{\qq\gl}}{s}$. 
Here I also observe that, as discussed in \sec{sec:Pi_and_QPH}, $\Im\[\Pi_{\qq\gl}\(\go\)\]$ is not enough to calculate the quasi--phonon solution of the
Dyson equation. The real part is also needed and, thus, we need an approach that treats real and imaginary parts on the same level.

\subsection{Variational properties of the phonon self--energy \myref{22-2-24}}\lab{sec:variational}
The approach of \ocite{Calandra2010,Berges2023,Caldarelli2025} is in this direction, as the aim of these works is to  use the variational freedom of the
self--energy (real and imaginary part) to find an expression that minimizes the error introduced by the static screening. The general idea is borrowed from
DFPT that is, indeed, a variational theory based on the extension of Hohenberg--Kohn theorem to density
perturbations~\cite{Baroni1987,Gonze1995a,Gonze1995b}. The DFPT variational concept 
have been extended, in ~\ocite{Calandra2010,Berges2023,Caldarelli2025}, to the phonon 
self--energy by using its (only formal~\cite{Marini2024}) analogy with the DFPT adiabatic dynamical matrix.

In the following I show how dynamical corrections emerge naturally from the variational 
analysis of $\Pi_{\qq\gl}\(\go\)$, in the same spirit of \ocite{Calandra2010,Berges2023,Caldarelli2025}.
I start by noticing that it is possible to define the  phonon--induced perturbation of the electronic density, $\gD n_{\qq\gl}\(\go\)$
\eql{eq:VAR.2}
{
 \amend{n_{\qq\gl}\(\go\)}=\evalat{n_{\qq\gl}\(\go\)}{e}+\evalat{n_{\qq\gl}\(\go\)}{p}.
}
In \e{eq:VAR.2} $\evalat{n_{\qq}\(\go\)}{e}$ is the purely electronic density and $\evalat{ n_{\qq\gl}\(\go\)}{p}$ its modification induced by the atomic perturbation
caused by an atomic shift along the normal mode $\(\qq\gl\)$. From
linear--response we know that, if the e--p interaction is weak, we can approximate 
\eql{eq:VAR.3}{
 \evalat{ n_{\qq\gl}}{p}\(\go\) \approx \chi^{0}_{\qq}\(\go\) \callG_{\qq\gl}\(\go\)=\chi_{\qq}\(\go\) g_{\qq\gl}.
}

If we use \e{eq:VAR.3} in \e{eq:VAR.1} we get
\eql{eq:VAR.4}
{
 \Pi_{\qq\gl}\(\go\)=  g_{\qq\gl}  \evalat{n_{\qq\gl}\(\go\)}{p}.
}
From \e{eq:VAR.4} it easily follows that, within linear--response,
\eql{eq:VAR.5}
{
\frac{\gd \Pi_{\qq\gl}\(\go\)}{\gd \evalat{n_{\qq^\p\gl^\p}\(\go\)}{p}}=g_{\qq\gl}\gd_{\qq\qq^\p}\gd_{\gl\gl^\p}.
}
From \e{eq:VAR.5} it is evident that, under the constrain given by \e{eq:VAR.3} the linear variation of the self--energy with respect to the density variation
is not zero. Keeping this in mind  \ocite{Calandra2010,Berges2023,Caldarelli2025} consider a more general functional of the $\evalat{ n_{\qq\gl}}{p}\(\go\)$ 
and $\chi^{0}_{\qq}\(\go\)$ obtained by observing that
\mll{eq:CCA.1}
{
\callG_{\qq\gl}\(\go\)=
\gee^{-1}_{\qq}\(\go\)g_{\qq\gl}=
g_{\qq\gl}+ V_\qq\chi_\qq\(\go\)g_{\qq\gl}=\\
g_{\qq\gl}+ V_\qq\chi^{0}_\qq\(\go\)\callG_{\qq\gl}\(\go\)=
g_{\qq\gl}+ V_\qq  \evalat{n_{\qq\gl}\(\go\)}{p}.
}
Let's now add and remove $\evalat{n_{\qq\gl}\(\go\)}{p} V_\qq$ to \e{eq:VAR.1}
\mll{eq:CCA.2}
{
\Pi_{\qq\gl}\(\go\)= 
\callG_{\qq\gl}\(\go\) \chi^{0}_\qq\(\go\) \callG_{\qq\gl}\(\go\)+\\
               -\evalat{n_{\qq\gl}\(\go\)}{p}V_\qq \evalat{n_{\qq\gl}\(\go\)}{p}.
}
\begin{widetext}
At this point \e{eq:CCA.1} is applied again on the first term on the r.h.s. of \e{eq:CCA.2}
\eql{eq:CCA.3}
{
\Pi_{\qq\gl}\(\go\)= \[g_{\qq\gl}+\evalat{n_{\qq\gl}\(\go\)}{p} V_\qq\] \chi^{0}_\qq\(\go\) 
\[g_{\qq\gl}+V_\qq\evalat{ n_{\qq\gl}\(\go\)}{p}\]
 -\evalat{ n_{\qq\gl}\(\go\)}{p}V_\qq \evalat{ n_{\qq\gl}\(\go\)}{p}.
}
\ocite{Calandra2010,Berges2023,Caldarelli2025} introduce, now, a new functional $\callF_{\qq\gl}\[\gr,\chi^0\]\(\go\)$
\eql{eq:CCA.3.1}
{
\callF_{\qq\gl}\[\gr,\chi^0\]\(\go\)= \[g_{\qq\gl}+\gr\(\go\) V_\qq\] \chi^{0}_\qq\(\go\) 
\[g_{\qq\gl}+V_\qq \gr\(\go\)\]-\gr\(\go\)V_\qq \gr\(\go\),
}
where, now, $\gr\(\go\)\neq \evalat{n_{\qq\gl}\(\go\)}{p}$ and, thus, it does not respect \e{eq:VAR.3}.  It follows the $\callF$ 
is a  functional of $\gr\(\go\)$ and, independently, of $\chi^{0}_\qq\(\go\)$. It easily follows that
\eqgl{eq:CCA.4}
{
\frac{1}{2}\evalat{\frac{\gd \callF_{\qq\gl}\[\gr,\chi^0\]\(\go\)}{\gd \gr\(\go\)}}{\gr\(\go\)=\evalat{n_{\qq\gl}\(\go\)}{p}}=
\evalat{V_\qq \[\chi^{0}_\qq\(\go\) \callG_\qq\(\go\)-\gr\(\go\)\]}{\gr\(\go\)=\evalat{n_{\qq\gl}\(\go\)}{p}}=0,\\
\frac{1}{2}\frac{\gd^2 \callF_{\qq\gl}\[\gr,\chi^0\]\(\go\)}{\gd \gr^2\(\go\)}=V_\qq\[\chi^{0}_\qq\(\go\)V_\qq-1\].
}
From \e{eq:CCA.4} it follows that, if we evaluate \e{eq:CCA.3.1} at $\gr\(\go\)\sim \evalat{ n_{\qq\gl}\(\go=0\)}{p}$ we get
\eql{eq:CCA.5}
{
\callF_{\qq\gl}\[\gr,\chi^0\]\(\go\)\sim \evalat{\Pi_{\qq\gl}\(\go\)}{SS}+
V_\qq\[\chi^{0}_\qq\(\go\)V_\qq-1\]\[\gr\(\go\)-\evalat{n_{\qq\gl}\(\go=0\)}{p}\]^2.
}
\end{widetext}
In \ocite{Calandra2010,Berges2023,Caldarelli2025} \e{eq:CCA.5} is used to state that, indeed, if we take a static approximation for the 
phonon induced density the error we introduce in the self--energy is second order in the density error. In \e{eq:CCA.5} we have introduced the
doubly statically--screened self--energy:
\eql{eq:CCA.6}
{
 \evalat{\Pi_{\qq\gl}\(\go\)}{SS}=g_{\qq\gl}\evalat{\callG_{\qq\gl}}{s}\evalat{\chi^{0}_{\qq}}{s}+\(\evalat{\callG_{\qq\gl}}{s}\)^2 \gD\chi^{0}_{\qq}\(\go\).
}
The coefficient of the second order in \e{eq:CCA.5} can be rewritten as 
\eql{eq:CCA.7}
{
 \frac{1}{2}\frac{\gd^2 \callF_{\qq\gl}\[\gr,\chi^0\]\(\go\)}{\gd \gr^2\(\go\)}=-V_\qq \gee_\qq\(\go\),
}
where $\gee_\qq\(\go\)=1-V_\qq\chi^{0}_\qq\(\go\)$. As $\chi^{0}_\qq\(\go\)\sim q^2$ and $V_\qq\sim\frac{1}{q^2}$ it follows that
$\lim_{\qq\rar\zero}\gee_\qq\(\go\)=O\(1\)$ and, consequently 
\eql{eq:CCA.8}
{
 \lim_{\qq\rar\zero}\frac{\gd^2 \callF_{\qq\gl}\[\gr,\chi^0\]\(\go\)}{\gd \gr^2\(\go\)}=\infty.
}
We can further work out \elab{eq:CCA.4}{b} in order to better evaluate the coefficient of the divergence appearing in \e{eq:CCA.8}.
Indeed we observe that (see \e{eq:JEL.5}) $V_\qq\sim \frac{g^2_{\qq\gl}}{\go_{\qq\gl}}$. This implies
\eql{eq:CCA.9}{
 \frac{\gd^2 \callF_{\qq\gl}\[\gr,\chi^0\]\(\go\)}{\gd \gr^2\(\go\)}\sim \frac{4\pi}{\go_{\qq\gl}q^2}\evalat{\Pi_{\qq\gl}\(\go\)}{BB}.
}
In \e{eq:CCA.9} I have introduced the bare--bare self--energy
\eql{eq:CCA.10}
{
 \evalat{\Pi_{\qq\gl}\(\go\)}{BB}=g^2_{\qq\gl} \chi^{0}_{\qq}\(\go\).
}
In general the screening (even if adiabatic) reduces the strength of $g_{\qq\gl}$. This means that we expect 
$\evalat{\Pi_{\qq\gl}\(\go\)}{BB}\gg \evalat{\Pi_{\qq\gl}\(\go\)}{SS}$.

\es{eq:CCA.8}{eq:CCA.9} demonstrate that, within the variational approach prposed by \ocite{Calandra2010,Berges2023,Caldarelli2025}, the second order coefficient
diverges when $\qq\rar\zero$ with a prefactor larger of the 0$^{th}$ order.
In practice, therefore, \e{eq:CCA.5} cannot justify {\em a priori} the use of a doubly statically--screened self--energy. We need a more
reliable and accurate criterion that I am going to introduce in the next sections.

\section{A dynamical perturbative expansion\myref{13-5-24}}\lab{sec:G_dyn}
\amend{In \ocite{Falter1988,Falter1981,Falter1980} C.Falter and collaborators have introduced a general renormalization theory of the electronic density reponse 
that allows to separate the dielectric function into a part which contains the relevant degrees of freedom plus a renormalizing term. Here I follow the same
strategy to derive a perturbative expansion of $\mat{\chi}^0_\qq\(\go\)$ around its static limit}.

In order to proceed we start from \e{eq:eq.6} written in the reference basis $\II_i\equiv\(n_i,m_i,\kk_i\)$. We split the electronic response function as
\eql{eq:PDE}
{
\mat{\chi}^0_\qq\(\go\)=\evalat{\mat{\chi}^0_\qq}{s}+\gD\mat{\chi}^0\(\go\).
}
In \e{eq:PDE} $\evalat{\mat{\chi}_\qq}{s}=\mat{\chi}_\qq\(t_2=t_1+0^+\)$ is the adiabatic electronic response function. We use, now, \e{eq:PDE} to expand the 
inverse dielectric matrix based on the identity~\footnote{See the notes of S.L.\,Adler about 
\href{https://www.ias.edu/sites/default/files/sns/files/1-matrixlog_tex(1).pdf}{\em Taylor Expansion and Derivative Formulas for Matrix Logarithms}.}
\eql{eq:VDE.2}
{
\[\mat{A}-\mat{B}\]^{-1}= \mat{A}^{-1}+\mat{A}^{-1}\mat{B}\[\mat{A}-\mat{B}\]^{-1}.
}
Thanks to \e{eq:VDE.2} we get
\mll{eq:VDE.3}
{
\mat{\gee}^{-1}_\qq\(\go\)= \evalat{ \mat{\gee}^{-1}_\qq}{s}+\evalat{ \mat{\gee}^{-1}_\qq}{s}\mat{V}_\qq\gD\mat{\chi}^0\(\go\)\mat{\gee}^{-1}_\qq\(\go\)=\\
 \evalat{ \mat{\gee}^{-1}_\qq}{s}+
\mat{\gee}^{-1}_\qq\(\go\)
\gD\mat{\chi}^0\(\go\)\mat{V}_\qq
\evalat{ \mat{\gee}^{-1}_\qq}{s}.
}
We can use now \e{eq:VDE.3} to derive an equation of motion for $\ul{\callG}_{\pm\qq}$ starting from $\evalat{\ul{\callG}}{s}$. Indeed from \e{eq:VDE.3}
and \e{eq:eq.7} it follows
\eql{eq:VDE.4}
{
\[\mat{1}-\evalat{ \mat{\gee}^{-1}_\qq}{s}\mat{V}_\qq\Delta\mat{\chi}^0_\qq\(\go\)\]\ul{\callG}_{\pm\qq\gl}\(\go\)=\evalat{\ul{\callG}_{\pm\qq\gl}}{s},
}
\e{eq:VDE.4} demonstrates that it is possible to write a formal relation between the screened electron--phonon vertex and its
statically screened limit written. We can now define  the dynamical electron--phonon vertex matrix $\mat{\gC}_{\qq\gl}\(\go\)$
so that 
\eql{eq:VDE.4.1}{
 \ul{\callG}_{\pm\qq\gl}\(\go\)=\mat{\gC}_{\qq\gl}\(\go\)\evalat{\ul{\callG}_{\pm\qq\gl}}{s},
}
with
\eql{eq:VDE.4.2}{
 \mat{\gC}_{\qq\gl}\(\go\)=
 \[\mat{1}-\evalat{ \mat{\gee}^{-1}_\qq}{s}\mat{V}_\qq\Delta\mat{\chi}^0_\qq\(\go\)\]^{-1}.
}
\e{eq:VDE.4.2} admits a {\bf formal} Taylor expansion
\eql{eq:VDE.4.3}{
 \mat{\gC}^n_{\qq\gl}\(\go\)=
 \sum_{m=0}^{n} \[ \evalat{\mat{\gee}^{-1}_\qq}{s}\mat{V}_\qq\Delta\mat{\chi}^0_\qq\(\go\)\]^m,
}
with $\mat{\gC}_{\qq\gl}\(\go\)=\lim_{n\rar \infty}  \mat{\gC}^n_{\qq\gl}\(\go\)$. 
\e{eq:VDE.4.3} is formally exact but meaningful only if the expansion terms are small enough. In practice this means that
it is possible to replace the full dynamical dependence with a statically screened interaction if and only if
\eql{eq:VDE.5}
{
\evalat{ \mat{\gee}^{-1}_\qq}{s}\mat{V}_\qq\Delta\mat{\chi}^0_\qq\(\go\)\ll \mat{1}.
}
\e{eq:VDE.5} define the necessary conditions to use a statically screened electron--phonon interaction.

We can now use \e{eq:VDE.4.3} to define the $n$--th order approximation to the phonon self--energy:
\seql{eq:VDE.8}{
\ml
{
 \orar{\Pi}^n_{\qq\gl}\(\go\)=
 \ul{g}_{-\qq\gl}^T 
 \bpg{\evalat{\mat{\chi}^0_\qq}{s} \mat{\gC}^n_{\qq\gl}\(\go\)+\\+
 \gD\mat{\chi}^0_\qq\(\go\) \mat{\gC}^{n-1}_{\qq\gl}\(\go\)} \evalat{\ul{\callG}_{\qq\gl}}{s},
}
and
\ml
{
 \olar{\Pi}^n_{\qq\gl}\(\go\)=
 \evalat{\ul{\callG}_{-\qq\gl}}{s}^T
 \bpg{
  \[\mat{\gC}^n_{\qq\gl}\(\go\)\]^T \evalat{\mat{\chi}^0_\qq}{s} +\\+
  \[\mat{\gC}^{n-1}_{\qq\gl}\(\go\)\]^T \gD\mat{\chi}^0_\qq\(\go\) } \ul{g}_{\qq\gl}.
}}
Thanks to \e{eq:VDE.8} we can rewrite \e{eq:eq.6} as
\eql{eq:VDE.9}
{
 \Pi_{\qq\gl}\(\go\)=\lim_{n\rar\infty}\frac{1}{2}\[ \orar{\Pi}^n_{\qq\gl}\(\go\)+\olar{\Pi}^n_{\qq\gl}\(\go\)\].
}

\subsection{Adiabatic limit and the doubly statically screened approximation}\lab{sec:Pi_0_and_1_order}
From \es{eq:VDE.4.1}{eq:VDE.4.3} it follows that
\eql{eq:PDE.1}
{
 \ul{\callG}^{n=0}_{\pm\qq\gl}\(\go\)= \mat{\gC}^{n=0}_{\qq\gl}\(\go\)\evalat{\ul{\callG}_{\pm\qq\gl}}{s}=\evalat{\ul{\callG}_{\pm\qq\gl}}{s}.
}
This means that at lowest order in the expansion, \elab{eq:VDE.8}{a} reduces to
\eql{eq:PDE.2}
{
 \orar{\Pi}^{n=0}_{\qq\gl}=
 \[\ul{g}_{-\qq\gl}\]^T
 \evalat{\mat{\chi}^{0}_{\qq}}{s}
 \evalat{\ul{\callG}_{\qq}}{s}=\olar{\Pi}^{n=0}_{\qq\gl}=\evalat{\Pi_{\qq\gl}}{s}.
}
We see that \e{eq:PDE.2} corresponds to the adiabatic and static limit of the phonon self--energy. 

If we move now to the 1$^{st}$ order in  $\Delta\mat{\chi}_\qq^0\(\go\)$ we obtain
\mll{eq:PDE.3}
{
 \orar{\Pi}^{n=1}_{\qq\gl}\(\go\)=\evalat{\Pi_{\qq\gl}}{s}+\\+
 \ul{g}^T_{-\qq\gl}\[ \evalat{\mat{\chi}^{0}_{\qq}}{s} \evalat{\mat{\gee}^{-1}_\qq}{s} \mat{V}_\qq+\mat{1}\]
 \Delta\mat{\chi}^{0}_{\qq}\(\go\) \evalat{\mat{\gee}^{-1}_\qq}{s}\amend{\ul{g}_{\qq\gl}}
}
Now we observe that 
\eql{eq:PDE.4}
{
\[\mat{1}+\evalat{\mat{\chi}^{0}_{\qq}}{s}
 \evalat{\mat{\gee}^{-1}_\qq}{s}
\mat{V}_\qq\]=\evalat{\mat{\gee}^{-1}_\qq}{s}.
}
\e{eq:PDE.4} can be verified order by order by expanding $\evalat{\mat{\gee}^{-1}_\qq}{s}$ in powers of $\mat{V}_\qq$. If we now use \e{eq:PDE.4} in \e{eq:PDE.3}
and use (see \e{eq:eq.2.1}) $\evalat{\mat{\gee}^{-1}_\qq}{s}=\[\evalat{\mat{\gee}^{-1}_\qq}{s}\]^T$, we finally get
\mll{eq:PDE.5}
{
 \orar{\Pi}^{n=1}_{\qq\gl}\(\go\)=\evalat{\Pi_{\qq\gl}}{s}+
 \[\evalat{\ul{\callG}_{-\qq\gl}}{s}\]^{T}
 \Delta\mat{\chi}^{0}_{\qq}\(\go\)
 \evalat{\ul{\callG}_{\qq\gl}}{s}=\\=\evalat{ \Pi_{\qq\gl}\(\go\)}{SS}.
}
\e{eq:PDE.5} demonstrates  that the statically double screened
phonon self--energy corresponds to the first order in the Taylor expansion of $\orar{\Pi}_{\qq\gl}\(\go\)$ in powers of 
$\Delta\mat{\chi}^0_\qq\(\go\)$. 

\e{eq:PDE.5} represents an alternative derivation  of \e{eq:CCA.5} with the crucial difference that \e{eq:VDE.8} allows to calculate exactly all corrective
terms that, in \e{eq:CCA.5}, are embodied in, the potentially diverging, second order term. This divergence appears now explictly as the case where the
formal Taylor expansion that defines $\olar{\orar{\Pi}}^{n}_{\qq\gl}\(\go\)$ does not converge. 

What we need now is to evaluate the corrections to \e{eq:PDE.5}, via \e{eq:VDE.8}, to see if they are important and under which conditions.

\section{The three--dimensional homogeneous electron gas}\lab{sec:jells}
\e{eq:VDE.8} provides  a formal and valid way to evaluate the validity of the statically screened approximation by calculating in practice the different orders
of the perturbative expansion. As this is very difficult to do in realistic materials and, in order to identify some general physical aspects of the problem, I
consider here the case of an exactly solvable model. This will allow us to estimate the importance of the perturbative terms of \e{eq:VDE.8}. The model is based
on the well known homogeneous electron gas\,(HEG). The HEG  is used a basic tool in several textbooks~\cite{Mahan1990,ALEXANDERL.FETTER1971} and works.  The
reason is that it is possible to derive exact properties that can provide valuable insights in the physics of the electron--phonon problem.

The definition of phonons and electron--phonon interaction in the HEG is complicated by the fact that, commonly, the atoms are replaced with a positively
charged jell of atoms~\cite{ALEXANDERL.FETTER1971}. In this case phonons are introduced as sound waves of the  ionic plasma whose
whose modes have a vanishing energy when $\qq\rar\zero$. However in solids it is the static screening of the dynamical matrix 
that produces vanishing phonon frequencies in the zero momentum limit. Thus a proper definition of phonons in the HEG must be derived by starting from a periodic array of atoms that, if the electrons are
assumed to be uniformly distributed, will produce just a single, bare, Drude--like atomic mode. 

\begin{figure}[t!]
{ \centering
\includegraphics[width=\columnwidth]{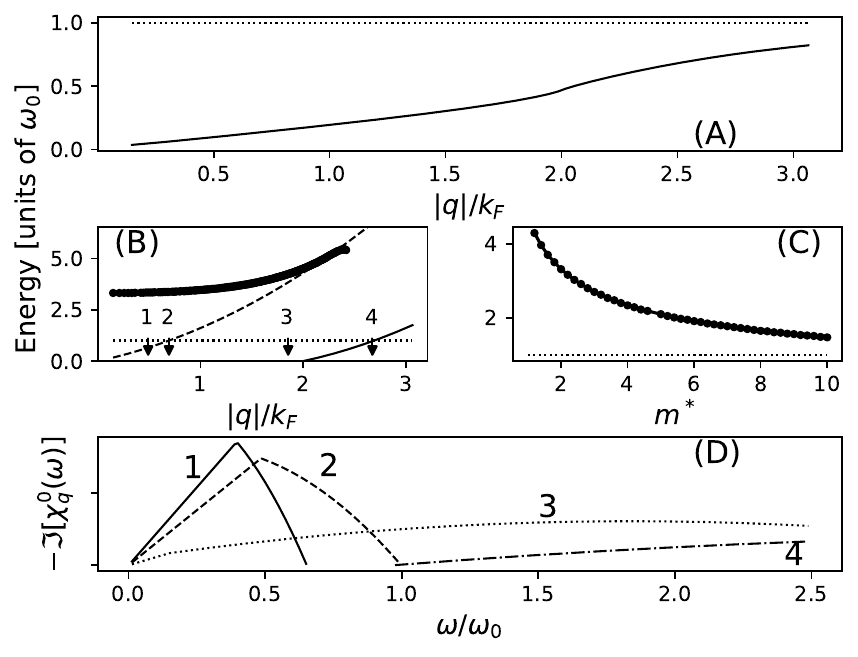} 
}
\caption{
\footnotesize{
In the frame\,A the Drude--like phonon energy $\go_0$, represented by an horizontal dotted line,
is shown together with the statically renormalized, acoustic, frequency (full line). In frame\,B the plasma energy
$\gO^{drude}_q$\,($\bullet\,\bullet\,\bullet$) is reported  as a function of the momentum $q$\,(in units of $k_F$)
together with $\go_0$\,(dotted line) and the two functions $E_{i=1,2}\(q\)$ (dashed and continuous lines) that define the lower and upper energy limits of the $\Im\[\Pi_q\(\go\)\]$. 
In frame\,C $\gO^{drude}_q$\,($\bullet\,\bullet\,\bullet$) is showed as a function of the electronic effective
mass $m^*$. Finally in frame\,D $\Im \[\chi^0_q\(\go\)\]$ is shown for the four representative values of $q$ indicated in frame\,B in the case of $m^*=2$.
}}
\label{fig:2}
\end{figure}

The introduction of phonons via a suitable lattice of atoms can be found in \ocite{Mahan1990,Schrieffer1999} and it has been recently
reviewed in \ocite{Stefanucci2023}. It is possible to extend it to the homogeneous case by approximating the electronic part of the Hamiltonian with an HEG. 
I will refer to this model as periodic homogeneous electron gas\,(PHEG). The form of the PHEG Hamiltonian can be  
formally derived from the mathematical scheme introduced in \sec{sec:H} by assuming:
\eqgl{eq:JEL.0}
{
 \mat{M}_{\qq}=\frac{1}{V_0}\frac{\qq\qq}{|\qq|^2},\\
 u_{n\kk}\(\xx\)=\frac{1}{\sqrt{V_0}}e^{\im\kk\cdot\xx}.
}
From \e{eq:JEL.0} it follows
\eqgl{eq:JEL.1}
{
 \bgc_{\qq}=\frac{\qq}{|\qq|},\\
 \go_{\qq}=Z\sqrt{\frac{4\pi}{V_0M}}=\go_0.
}
Thanks to \es{eq:JEL.0}{eq:JEL.1} the PHEG Hamiltonian can be derived from \es{eq:H.3}{eq:H.12} and shown to be
\mll{eq:JEL.2}
{
 \h{H}=\sum_\kk \gee_\kk \h{c}^\dag_{\kk}\h{c}_{\kk}+
 \frac{4\pi}{2V_c}\sum_\qq \frac{ \h{\gr}_{\qq} \h{\gr}_{-\qq}}{q^2}+\go_0\sum_\qq\h{n}_\qq+\\+
 \frac{1}{\sqrt{V_c}} \sum_\qq g_q \h{\gr}_{-\qq}\frac{\(\h{b}_\qq+\h{b}^\dag_{-\qq}\)}{\sqrt{2}}.
}
In \e{eq:JEL.2} $C^{ref}_q$  (see \e{eq:DP.4.2}) is zero by definition and also the Ehrenfest, tad--pole diagram is regularized to zero as discussed 
in~\cite{Schrieffer1999,ALEXANDERL.FETTER1971}. Moreover
\eql{eq:JEL.3}
{
 g_q=\frac{Z}{\sqrt{V_0 M\go_0}}\frac{4\pi}{q}=\frac{\sqrt{4\pi \go_0}}{q},
}
and
\eql{eq:JEL.4}
{
 \h\gr_\qq=\sum_\kk \h{c}^\dag_{\kk-\qq}\h{c}_{\kk}.
}
\e{eq:JEL.2}
is rotationally invariant. As a consequence the independent--particle response function depends only on the momentum modulus $q=\phmod{\qq}$.
The same applies to the all $\qq$--dependent quantities.
$g_q$ has the peculiar property~\cite{Schrieffer1999}
\eql{eq:JEL.5}
{
 V_q=\frac{g_q^2}{\go_0}.
}
Thanks to \e{eq:JEL.5} it is simple algebra to demonstrate~\cite{Schrieffer1999} that the static renormalization of the bare, Drude like, phonon mode produces
the acoustic branch
\eql{eq:JEL.6}{
 \(\evalat{\gO_q}{s}\)^2=\go_0^2+\go_0\evalat{\Pi_q}{s}=\evalat{\gee_q^{-1}}{s}\go_0^2\xrightarrow[q\rar 0]{} q^2.
}
\e{eq:JEL.6} is numerically verified in \figlab{fig:2}{A}. 

$\chi^0_q\(\go\)$ can be calculated exactly (see \ocite{ALEXANDERL.FETTER1971}) and the electronic response of the system is entirely defined
by specifying the effective mass $m^*$ and the Fermi momentum $k_F$. Equivalently one can use the electronic density $n_{el}=\frac{K_F}{3\pi^2}$. $k_F$ is
evaluated numerically by specifying the number of electrons per unit cell.

The HEG response function is dominated by an isolated pole, the Drude plasmon, with energy $\gO^{drude}_q$\cite{ALEXANDERL.FETTER1971}:
\eql{eq:JEL.10}{
 \gO^{drude}_q=\sqrt{4\pi n_{el}}\[1+\ga q^2+\dots\].
}
The calculated $\gO^{drude}_q$ is shown in \figlab{fig:2}{B} as a function of $q$ and in \figlab{fig:2}{C} as a function of the
effective mass $m^*$.

The HEG response function main feature is that
\eql{Eq:JEL.11}
{
 \Im\[\chi^0_q\(\go\)\]\neq 0  \quad \go\in\[E_1\(q\),E_2\(q\)\].
}
The functions $E_{1}\(q\)$\,($E_2\(q\)$) are shown in \figlab{fig:2}{B} as continuous\,(dashed) line. 

In all numerical calculations I considered $n_{el}=0.2$. This specific value does not affect the
physical conclusions as the HEG properties can be tuned, as shown in \ocite{ALEXANDERL.FETTER1971}, by using the variable  $\frac{q}{m^*}$.
I will, instead considered different values of $m^*$ to span different values of $\gO^{drude}_q$ (see \figlab{fig:2}{C}).

\subsection{On--the--mass shell versus quasi--phonon approximation}
\begin{figure}[H]
{ \centering
\includegraphics[width=\columnwidth]{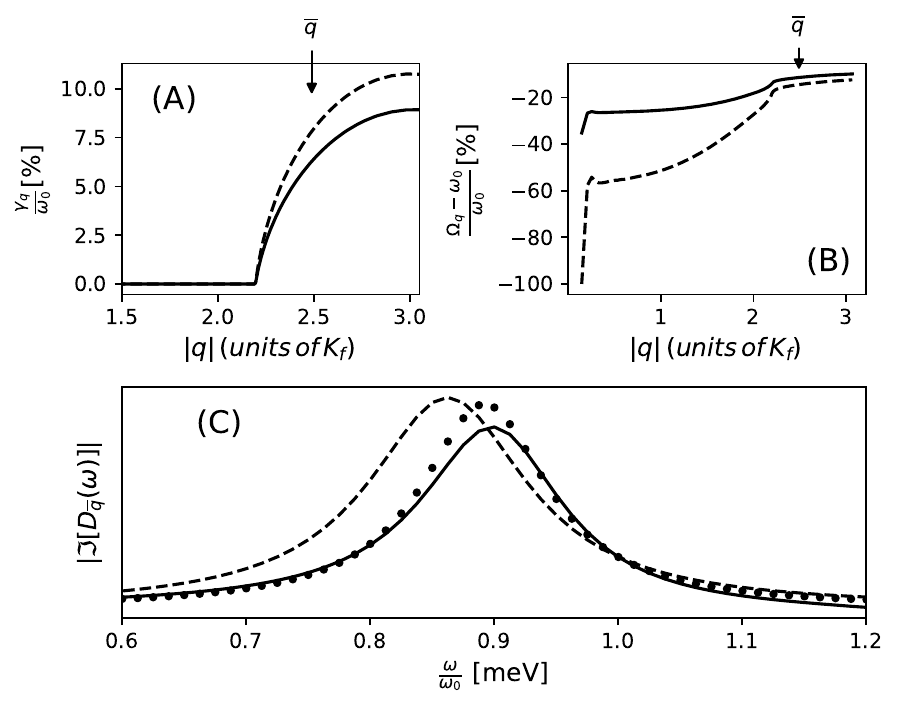} 
}
\caption{
\footnotesize{
Upper left frame: QPH\,(continuous line) versus OMS\,(dashed line) phonon width as a function of the phonon momentum (in units of the Fermi energy, $k_F$). 
In the two lower frames I compare the exact phonon spectral function ($\bullet\,\bullet\,\bullet$) with the one evaluated within the OMS\,(dashed line)
and QPH\,(continuous line) in the case of $q=\oo{q}$. In this case $m^*=10$.}
}
\label{fig:3}
\end{figure}

\lab{sec:JELL_QPH}
In the PHEG the phonon self--energy acquires a simple form
\eql{eq:JQPH.1}{
 \Pi_q\(\go\) = g_{q}^2 \chi_q\(\go\)=g_{q} \chi^0_q\(\go\) \callG_q\(\go\).
}
Here I use the time--dependent Hartree approximation for $\chi_q\(\go\)$
\eql{eq:JQPH.1.1}
{
 \chi_q\(\go\)=\frac{ \chi^0_q\(\go\) }{1-\frac{4\pi}{q^2}  \chi^0_q\(\go\)}.
}
\e{eq:JQPH.1} can be used  to verify the accuracy of the on--the--mass shell\,(OMS) approximation. This is done in \fig{fig:3} in the case of $m^*=5$. From
\figlab{fig:2}{C} this value of the effective electronic mass corresponds to $\gO^{plasma}_q\sim \go_0$, a case where we expect strong non--adiabatic effects.

In \figlab{fig:3}{A} and \figlab{fig:3}{B} the renormalized phonon width and energy are calculated both within the OMS\,(dashed line) and QPH\,(continuous line).
The OMS systematically overestimates both $\gc_q$ and $\gO_q$. This overestimation is particularly large in the low--energy regime where it can be as large as
twice the correction.

I now consider the specific $q=\oo{q}$ indicated in \fig{fig:3}.  From \figlab{fig:3}{A} we see that $\gamma^{QPH}_{\oo{q}}\approx 10\% \go_0$.
In \figlab{fig:3}{C} I compare the QPH\,(OMS) spectral functions
\eql{eq:JQPH.2}
{
 \evalat{D_{\oo{q}}\(\go\)}{kind}=\frac{1}{2}\sum_{s=\pm}\frac{s}{\go-s\gO^{kind}_{\oo{q}}+\im \gc^{kind}_{\oo{q}}},
}
where $kind=OMS,QPH$, with the exact one\,($\bullet\,\bullet\,\bullet$). The OMS spectral function deviates from the exact solution that, instead is well
described by the QPH approximation.

\subsection{Validation of the dynamical perturbative expansion}\lab{sec:JELL_Pi}
\begin{figure}[h!]
\begin{center}
\includegraphics[width=\columnwidth]{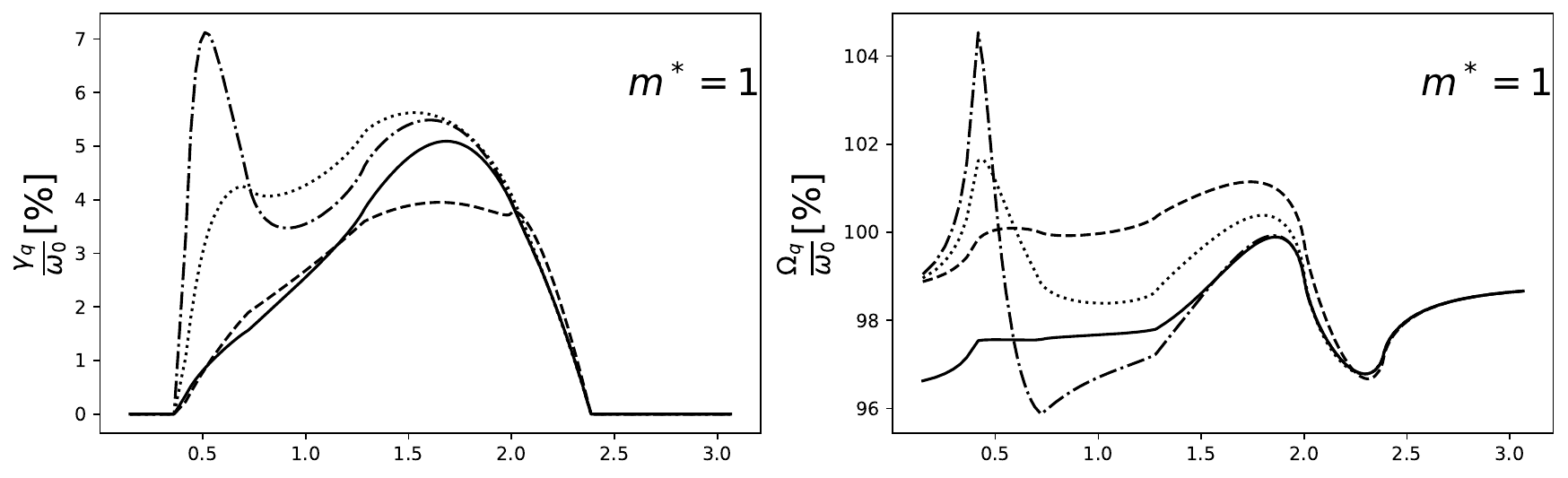} 
\\
\includegraphics[width=\columnwidth]{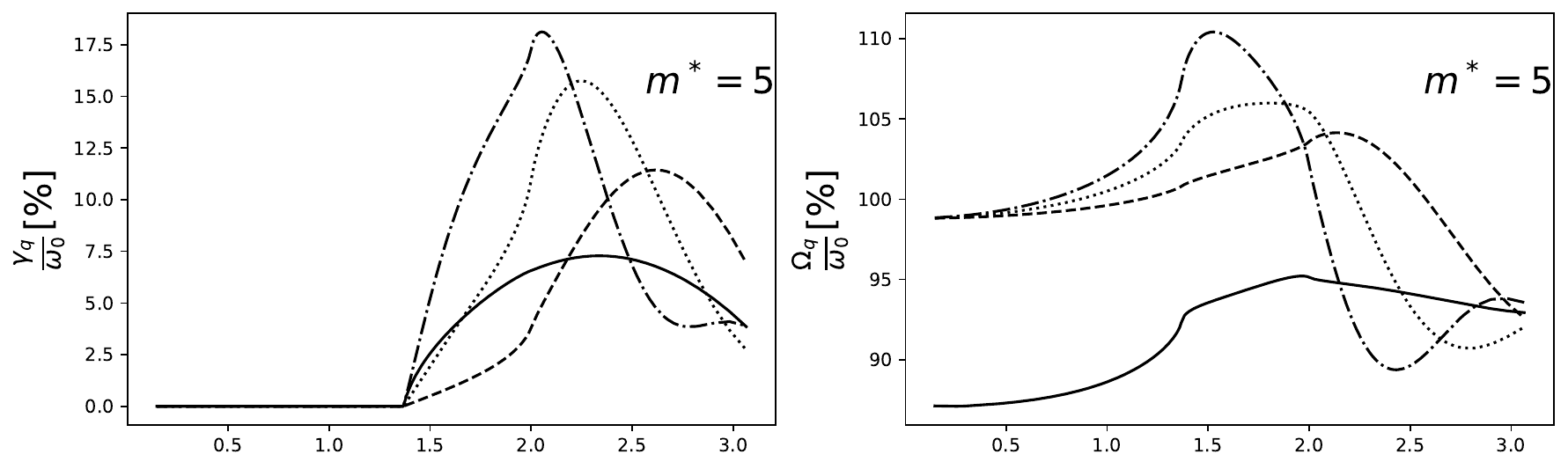} 
\\
\includegraphics[width=\columnwidth]{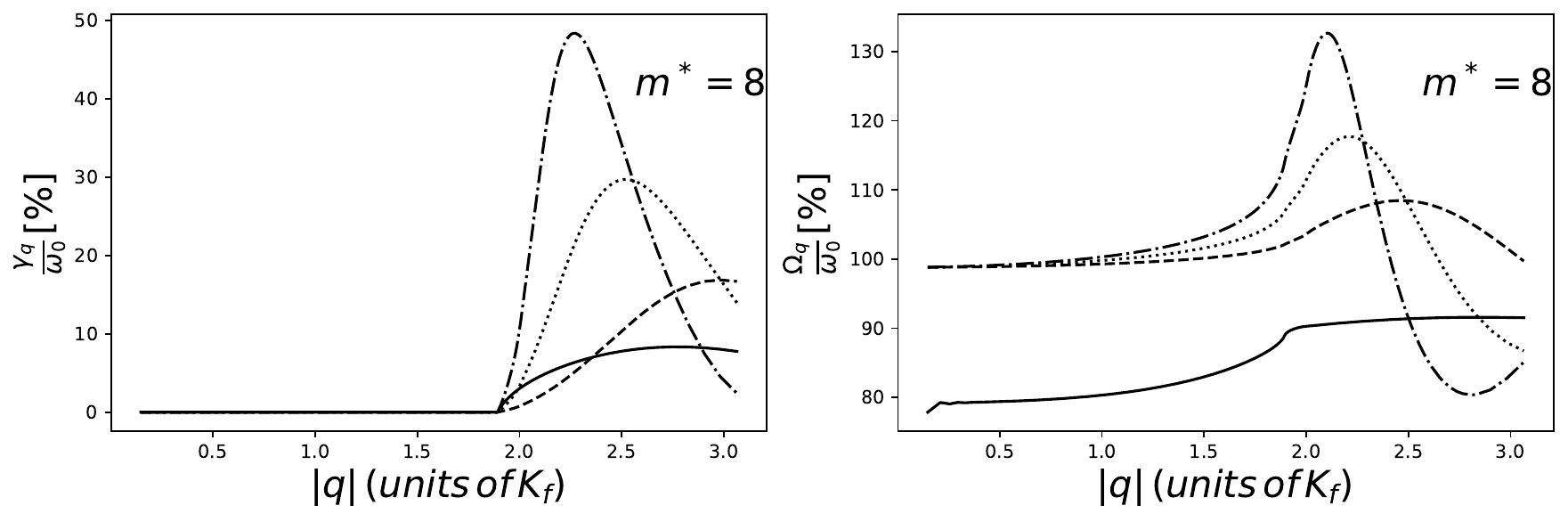} 
\end{center}
\caption{
\footnotesize{QPH energy, $\gO_q$, and width, $\gc_q$, calculated by using the exact self--energy\,(continuous line), and the $n \leq 3$ orders of
the perturbative expansion, \e{eq:JPI.4}. The PHEG considered here corresponds to $m^*=1$ (upper frames), $m^*=5$ (middle frames)
and $m^*=8$ (lower frames). 
The different orders are represented using the dashed\,($n=1$), dotted\,($n=2$) and dot--dashed\,($n=3$) lines. We see that in all cases the perturbative expansion is
not converging.
As expected from simple physical arguments the error of a given $n$ order 
increases as $m^*$ decreases (which corresponds to a decreasing plasma frequency). 
}}
\label{fig:4}
\end{figure}

In the PHEG \es{eq:VDE.8}{eq:VDE.9} acquire a simple form
\eql{eq:JPI.4}
{
 \Pi^n_q\(\go\)=g^{\remove{2}}_{q}\[ \evalat{\chi^0_q}{s}  \gC^n_q\(\go\)+\Delta\chi^0_q\(\go\) \gC^{n-1}_q\(\go\)\]\amend{\evalat{\callG_{\qq}}{s}}.
}
where
\eql{eq:JPI.5}
{
 \gC_q^n\(\go\)=\sum_{m=0}^n\[\frac{V_q\Delta\chi^0_q\(\go\)}{\evalat{\gee_q}{s}}\]^m,
}
and $\gC^0_q\(\go\)=1$.

In order to validate \e{eq:JPI.4} we can compare the finite orders, $\Pi^n_q\(\go\)$ with the $n\rar\infty$ limit corresponding to the exact self--energy
\eql{eq:JPI.6}
{
 \Pi_q\(\go\)=\lim_{n\rar\infty} \Pi^n_q\(\go\)=g_q \evalat{\callG_q}{s} \gC_q\(\go\) \chi^0_q\(\go\).
}
If we not notice that
\eql{eq:JPI.7}
{
 \evalat{\Pi_q\(\go\)}{BS}=g_q \evalat{\callG_q}{s} \chi^0_q\(\go\),
}
it also follows that
\eql{eq:JPI.7.1}
{
 \Pi_q\(\go\)= \gC_q\(\go\) \evalat{\Pi_q\(\go\)}{BS}.
}
It is important to note that, as demonstrated in \sec{sec:Pi_0_and_1_order}, the  doubly screened self--energy commonly used in the literature
corresponds to the $n=1$ order of the self--energy when expanded in powers of the dynamical term $\frac{V_q\Delta\chi_q\(\go\)}{\evalat{\gee^{-1}_q}{s}}$:
\eql{eq:JPI.8}
{
 \Pi^1_q\(\go\)=\evalat{\Pi_q\(\go\)}{SS}=\evalat{\Pi_q}{s}+\evalat{\callG_q}{s}^2 \gD\chi^0_q\(\go\).
}
In \fig{fig:4} the $n$--th order phonon self--energy is used to derive the corresponding $n$--th order $\go^{QPH}_q$ and $\gc^{QPH}_q$ that I show, in units of
$\go_0$ for three representative values of $m^*=1,5,8$. These values of the effective mass correspond to different ratio $\frac{\gO^{drude}_q}{\go_0}$, see
\figlab{fig:2}{C}. 

From \fig{fig:4} we see a common trend: in all cases the terms $\Pi^n_q\(\go\)$ does not decreases increasing the order $n$. This means that, in practice, the
Taylor expansion \e{eq:JPI.4} does not converge. What changes as a function of the effective mass is the absolute intensity of 
 $\Pi_q\(\go\)$ and, consequently, the intensity of its $n$--th Taylor expansion term. If we increase $m^*$ the situation worsens as the electronic plasma
frequency approaches the phonon energy.

The final message of \fig{fig:4} is that dynamical corrections cannot be treated perturbatively and that the $\evalat{\Pi_q\(\go\)}{SS}$ is not a reasonable
approximation. 

\subsection{The dynamical vertex function}\lab{sec:JELL_Gamma}
\begin{figure}[t!]
{ \centering
\includegraphics[width=\columnwidth]{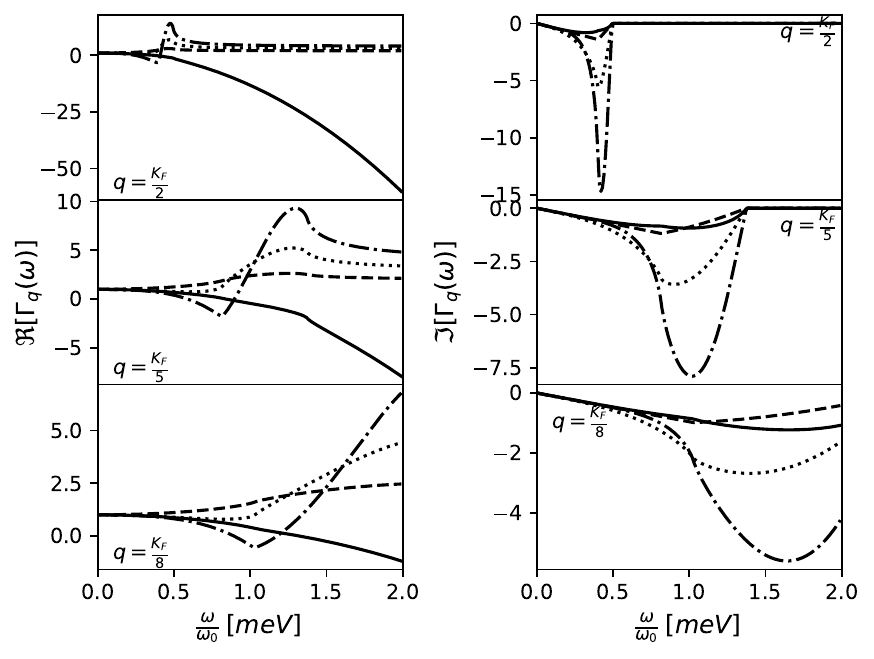} 
}
\caption{\footnotesize{
The $\gC^{n \leq 3}_q\(\go\)$ dynamical vertex function in the PHEG model corresponding to $m^*=5$, $n_{el}=0.2$ and $\go_0=400$\,meV.
The different orders are represented using the dashed\,($n=1$), dotted\,($n=2$) and dot--dashed\,($n=3$) lines. The non perturbative expression is
represented by the continuous line.
}}
\label{fig:5}
\end{figure}

In \fig{fig:5} I consider the $m^*=5$ case, which represents a moderate e--p interaction case\,(see the middle frames of \fig{fig:4}) as the
Drude plasma frequency is $\gO^{drude}_q\sim3\go_0$ (see \figlab{fig:2}{C}). We see that the presence of an isolated peak in the HEG screening function produces,
especially in the low momentum range, a very large vertex dynamical correction that cannot be described using the Taylor expansion.

The situation improves increasing $q$ as the plasma frequency moves away from the phonon energy. Still dynamical corrections cannot be neglected and the doubly
screened approximation ($n=1$ case) is still far from the exact solution.

In order to find a simple way to evaluate dynamical effects let's derive an exact relation between $\gC_q\(\go\)$ and $\evalat{\Pi_q\(\go\)}{SS/BS}$.
Let's start by \amend{defining
\eqgl{eq:JG.1}
{
\gC_q\(\go\)=\[1-\gd_q\(\go\)\]^{-1},\\
\gd_q\(\go\)=\evalat{\gee^{-1}_q}{s}V_q\Delta\chi^0_q\(\go\).
}
}
\remove{Moreover}  \e{eq:PDE.5} demonstrates that
$\evalat{\Pi_q\(\go\)}{SS}$  corresponds to the first order expansion in powers of $\gd_q\(\go\)$. Thus from \e{eq:JPI.6} it follows that
\eql{eq:JG.2}{
\evalat{\Pi_q\(\go\)}{SS}= \evalat{\Pi_q\(\go\)}{BS}+\gd_q\(\go\) \evalat{\Pi_q}{s}.
}
We can now invert \e{eq:JG.2} and obtain
\eql{eq:JG.3}{
 \gd_q\(\go\)=\frac{ \evalat{\Pi_q\(\go\)}{SS} - \evalat{\Pi_q\(\go\)}{BS}}{\evalat{\Pi_q}{s}}.
}
\e{eq:JG.3} demonstrates that larger is the difference between $\evalat{\Pi_q\(\go\)}{SS}$ and $\evalat{\Pi_q\(\go\)}{BS}$, larger is the  strength of the 
dynamical vertex function. It is important to remind that the \amend{bare--statically screened} limit corresponds to $\gd_q\(\go\)=0$, while the doubly statically screened self--energy
\amend{is valid only when} $\gd_q\(\go\)\amend{\gg} \[ \gd_q\(\go\) \]^2$.

It is clearly difficult, from \e{eq:JG.3}, to obtain a quantitative evaluation of the dynamical vertex corrections. A realistic calculation, indeed, can be provided only
by calculating the phonon and electronic self--energy with the fully screened e--p interaction. Still \e{eq:JG.3} can give a clear (and easy to calculate)
indication of the potential dynamical corrections. 

We conclude this section by further simplifing \e{eq:JG.3} by using the QPH picture, \amend{introduced} in \sec{sec:Pi_and_QPH}.
\remove{If we use \e{eq:QPH.2} to rewrite \e{eq:JG.3} in terms of $\gc_{\qq\gl}$ and $Z_{\qq\gl}$.}
\amend{We start observing that from \elab{eq:QPH.4}{a}--\elab{eq:QPH.4}{b} it follows}
\eqgl{eq:JG.4}{
 \Im\[\gb_{\qq\gl}\]=-\frac{2\gc_{\qq\gl}}{\go_{\qq\gl}Z_{\qq\gl}},\\
 \Re\[\gb_{\qq\gl}\]=1-\frac{1}{Z_{\qq\gl}}.
}
\amend{If we now use \e{eq:JG.4} into \elab{eq:QPH.2}{a} and transform $q\rar\qq\gl$ we get}
\mll{eq:JG.5}{
 \evalat{\Pi_{\qq\gl}\(\go\)}{kind}=\evalat{\Pi_{\qq\gl}}{s}-\im\frac{2\go}{\go_{\qq\gl}}\frac{\evalat{\gc_{\qq\gl}}{kind}}{\evalat{Z_{\qq\gl}}{kind}}+\\+
 \(1-\frac{1}{\evalat{Z_{\qq\gl}}{kind}}\)\frac{\go^2}{\go^2_{\qq\gl}},
}
with $kind=SS,BS$. From \e{eq:JG.5} it finally follows
\amend{
\mll{eq:JG.6}{
 \gd_{\qq\gl}\(\go\)=\frac{\go}{\evalat{\Pi_{\qq\gl}}{s} \go_{\qq\gl}}
 \[\go\( \frac{1}{\evalat{Z_{\qq\gl}}{SS}}-\frac{1}{\evalat{Z_{\qq\gl}}{BS}}\)+\nl+
 2\im\(\frac{\evalat{\gc_{\qq\gl}}{SS}}{\evalat{Z_{\qq\gl}}{SS}}-\frac{\evalat{\gc_{\qq\gl}}{BS}}{\evalat{Z_{\qq\gl}}{BS}}\)\].
}
We can further simplify \e{eq:JG.6} to obtain an estimate of $\gd_{\qq\gl}\(\go\)$ if we assume $|\gb_{\qq\gl}|\ll 1$. In this case, by using \e{eq:QPH.7}, we
get
\eqgl{eq:JG.7}
{
 Z_{\qq\gl}\sim 1+\Re\[\gb_{\qq\gl}\]=1+\frac{2 \(\gO^{OMS}_{\qq\gl}-\go_{\qq\gl}\)}{\go_{\qq\gl}},\\
 \frac{\gc_{\qq\gl}}{Z_{\qq\gl}}\sim \gc^{OMS}_{\qq\gl}.
}
From \e{eq:JG.7} we finally obtain
\eqgl{eq:JG.8}{
 \Re\[\gd_{\qq\gl}\(\go\)\]=\frac{\go^2}{2\go^2_{\qq\gl}\evalat{\Pi_{\qq\gl}}{s}}\(\evalat{\gO^{OMS}_{\qq\gl}}{SS}-\evalat{\gO^{OMS}_{\qq\gl}}{BS}\),\\
 \Im\[\gd_{\qq\gl}\(\go\)\]=\frac{2\go}{\go_{\qq\gl}\evalat{\Pi_{\qq\gl}}{s}}\(\evalat{\gc^{OMS}_{\qq\gl}}{SS}-\evalat{\gc^{OMS}_{\qq\gl}}{BS}\).
}
\e{eq:JG.8} is really instructive as it allows, by means of a simple on--the--mass shell approximated calculation of the phonon energy and width, to estimate the
vertex dynamical correction function. We can further manipulate \e{eq:JG.8} by assuming that $\Im\[\gd_{\qq\gl}\(\go=\go_{\qq\gl}\)\]\ll 1$ and approximating
$\evalat{\gO^{OMS}_{\qq\gl}}{BS}\sim \go_{\qq\gl}$ it follows that
\eql{eq:JG.9}{
 \Re\[\gd_{\qq\gl}\(\go_{\qq\gl}\)\]\sim \frac{\evalat{\gO^{OMS}_{\qq\gl}}{SS}-\go_{\qq\gl}}{2 \evalat{\Pi_{\qq\gl}}{s}}.
}
In \ocite{Saitta2008} Saitta et al. calculated the non--adiabatic corrections to the phonon frequency of several materials. From Table\,I of their work we
can see that $\(\evalat{\gO^{OMS}_{\qq\gl}}{SS}-\go_{\qq\gl}\)=\ga \go_{\qq\gl}$ with the prefactor $\ga$ that ranges from $\ga=0.06$ in KC$_{24}$ to $\ga=0.25$
in KC$_8$ and even $\ga=0.41$ in MgB$_2$\,(at $21K$ temperature). It follows that, within these approximations, $\Re\[\gd_{\qq\gl}\(\go_{\qq\gl}\)\]>0$, which
implies $\gC_{\qq\gl}\(\go_{\qq\gl}\)>1$. 
From \e{eq:JG.9} we see that $\Re\[\gd_{\qq\gl}\(\go_{\qq\gl}\)\]$ can be potentially large enough to make the
first order expansion of $\gC_{\qq\gl}\(\go_{\qq\gl}\)$ inaccurate.  However the quantitative deviation from a doubly statically screened approximation in
realistic materials requires an accurate evaluation of $\gC_{\qq\gl}\(\go\)$, that is beyond the scope of this work.
}

\section{Conclusions}\lab{sec:conclusions}
Several conclusions can be drawn from the results presented in this work.  Many years of literature, grounded on e--p models or indirect theoretical and
numerical results, have instilled the idea that the electron--phonon vertex can be replaced, in the Hamiltonian, with a statically screened expression.  This
assumption is currently used in most of the calculations of physical phenomena caused by the electron--phonon interaction and it is also coded in many \ai\,
codes.

In this work I have provided a mathematically accurate, and formally exact, way of including retardation effects via a dynamical vertex correction
$\gC_{e-p}\(\go\)$.  The properties of this vertex function have been discussed in general and calculated in practice in an exactly solvable periodic
homogeneous electron gas.  The model has been used to perform a detailed assessment of the performance of the different approximations corresponding to
different orders in the Taylor expansion of $\gC_{e-p}\(\go\)$.  The commonly used bare--screened and screened--screened approximations are obtained at the
lowest order of this perturbative expansion.  

The results presented here unequivocally demonstrate what could be expected on the basis of physical intuition: 
dynamical screening effects impose a combined description where the phonon scatters with a mixture of
single--particle and plasmonic excitations. If the energy of these excitations approaches the phonon frequency and, in general, the energy range spanned by the vertex
function, dynamical effects can be so large to be impossible to be described perturbatively. In this case any description based on statically screened vertexes
is bound to fail.

In order to have a tool to evaluate the importance of dynamical corrections,
I have proposed a simple form of $\gC_{e-p}\(\go\)$ written in terms of simple bare--screened and screened--screened calculations. Simulations that are possible
to do even \ai. This simplified vertex function can be used to evaluate the impact of the dynamical screening in  realistic materials.

\section{Acknowledgments}
A.M. acknowledges the funding received from: 
MaX {\em MAterials design at the eXascale}, a European Centre of Excellence funded by the European Union’s program HORIZON-EUROHPCJU-2021-COE-01 (Grant No.
101093374); {\em Nanoscience Foundries and Fine Analysis -- Europe | PILOT} H2020-INFRAIA-03-2020 (Grant No. 101007417); 
{\em PRIN: Progetti di Ricerca di rilevante interesse Nazionale} Bando 2020 (Prot. 2020JZ5N9M).

\appendix

\section{Equation of motion for the time--dependent Hartree screening \myref{28-3-4 EOM for $\chi$}}\lab{APP:TDH}
A crucial ingredient introduced in \sec{sec:vertex_and_SEs} is the inverse dielectric function. This has been introduced in \e{eq:DP.1}
and is defined in terms of the reducible density--density response function $\chi_\qq\(1,2\)$
\eql{eq:We.1}
{
 \gee^{-1}_\qq\(1,2\)=\gd\(1,2\)+\int \di 3 V_\qq\(1,3\)\chi_\qq\(3,2\).
}
In \e{eq:We.1} $\chi_\qq\(3,2\)$ can be written in the independent particle basis defined in \sec{sec:H}:
\begin{widetext}
\eql{eq:We.2}
{
 \chi_\qq\(1,2\)=\frac{1}{V_c}
 \sum_{n_1 m_1\kk_1}
 \sum_{n_2 m_2\kk_2}
 \oo{u_{n_1\kk_1-\qq}\(\xx_1\)} u_{m_1\kk_1}\(\xx_1\)
 \oo{u_{n_2\kk_2}\(\xx_2\)} u_{m_2\kk_2-\qq}\(\xx_2\)
 \chi_{\qq,\sst{n_1 m_1\kk_1}{n_2 m_2\kk_2}}\(z_1,z_2\).
}
In \e{eq:We.2} I have introduced the basis representation of the response function 
\mll{eq:We.3}
{
 \im \chi_{\qq,\sst{n_1 m_1\kk_1}{n_2 m_2\kk_2}}\(z_1,z_2\)=
 \average{ \callT_c\bpg{ \h{c}^\dag_{n_1\kk_1-\qq}\(z_1\) \h{c}_{m_1\kk_1}\(z_1\) \h{c}^\dag_{n_2\kk_2}\(z_2\) \h{c}_{m_2\kk_2-\qq}\(z_2\) } }+\\
 -\average{\h{c}^\dag_{n_1\kk_1-\qq}\(z_1\) \h{c}_{m_1\kk_1}\(z_1\)}\average{\h{c}^\dag_{n_2\kk_2}\(z_2\) \h{c}_{m_2\kk_2-\qq}\(z_2\)} .
}
\end{widetext}
In order to simplify the notation I introduce a combined index for bands pairs and momenta $\II_i\equiv\(n_i,m_i,\kk_i\)$
\eql{eq:We.4}
{
 \chi_{\qq,\sst{n_1 m_1\kk_1}{n_2 m_2\kk_2}}\(z_1,z_2\)\equiv \chi_{\qq\II_1\oo{\II}_2}\(z_1,z_2\).
}
The index $\oo{\II}_2$ in \e{eq:We.4} is indicating that the order of the two indexes is exchanged, as defined in \e{eq:We.3}.

Within the notation of \e{eq:We.4} the non--interacting case corresponds to
$\evalat{\chi_{\qq\II_1\oo{\II}_2}\(z_1,z_2\)}{IP}=\gd_{\II_1\II_2}\chi^0_{\qq\II_1}\(z_1,z_2\)$. Note that in the present case the irreducible response function corresponds to
the independent particle approximation.

I now consider the retarded component of the equilibrium response function, $\mat{\chi}_\qq\(t_1-t_2\)$. The corresponding equation of motion 
can be easily derived by using different strategies: it can be derived diagrammatically~\cite{Strinati1988} or
by using the corresponding equation of motion for the density matrix~\cite{Attaccalite2011b}. 
The final result is
\eql{eq:3.3.8}
{
\mat{\chi}_\qq\(\go\)=\mat{\chi}^{0}_\qq\(\go\)\[\mat{1}+\mat{V}_\qq\mat{\chi}_\qq\(\go\)\].
}
\e{eq:3.3.8} is the Dyson equation for $\chi$ corresponding to the time--dependent Hartree approximation, also known as Random Phase Approximation\,(RPA).

In \e{eq:3.3.8} we have that
\eql{eq:3.3.5}
{
\chi^{0}_{\qq\II_1}\(\go\)=-\frac{1}{V_c}\( \frac{f_{n_1\kk_1-\qq}\(\gb\)-f_{m_1\kk_1}\(\gb\)}{\go+\gee_{m_1\kk_1}-\gee_{n_1\kk_1-\qq}+i0^+}\).
}
In \e{eq:3.3.5} $0^+$ is a tiny positive number that will be sent to zero at the end of the calculation.

\FloatBarrier 

\bibliography{paper}

\end{document}